\documentclass[useAMS,usenatbib]{mnras}

\usepackage{url,times,hyperref,graphicx,amsmath,amsfonts,amssymb,aas_macros,xcolor,epsfig,epstopdf,ulem}
\usepackage{dblfloatfix} 

\newcommand{\mhalo}{M$_{\rm halo}$}

\newcommand{\msun}{M$_\odot$}

\newcommand{\hMpc}{{\ifmmode{h^{-1}{\rm Mpc}}\else{$h^{-1}$Mpc}\fi}}
\newcommand{\hkpc}{{\ifmmode{h^{-1}{\rm kpc}}\else{$h^{-1}$kpc}\fi}}
\newcommand{\hMsun}{{\ifmmode{h^{-1}{\rm {M_{\odot}}}}\else{$h^{-1}{\rm{M_{\odot}}}$}\fi}}
\newcommand{\ltsima}{$\; \buildrel < \over \sim \;$}
\newcommand{\gtsima}{$\; \buildrel > \over \sim \;$}
\newcommand{\lsim}{\lower.5ex\hbox{\ltsima}}
\newcommand{\gsim}{\lower.5ex\hbox{\gtsima}}

\def\lesssim{\mathrel{\hbox{\rlap{\hbox{\lower4pt\hbox{$\sim$}}}\hbox{$<$}}}}
\def\gtrsim{\mathrel{\hbox{\rlap{\hbox{\lower4pt\hbox{$\sim$}}}\hbox{$>$}}}}

\newcommand{\beq}{\begin{equation}}
\newcommand{\eeq}{\end{equation}}
\def\beqa{\begin{eqnarray}}
\def\eeqa{\end{eqnarray}}
\def\hMpc{$h^{-1}\,{\rm Mpc}$}
\def\hkpc{$h^{-1}\,{\rm kpc}$}

\def\head{
 \vbox to 0pt{\vss
                   \hbox to 0pt{\hskip 440pt\rm LA-UR-10-07069\hss}
                  \vskip 25pt}}

\title[Deep Learning to distinguish Cusps vs Cores]
{A probabilistic deep learning model to distinguish cusps and cores in dwarf galaxies}
\author[Exp\'{o}sito-M\'{a}rquez et al.]
      {J. Exp\'{o}sito-M\'{a}rquez$^{1,2}$, C. B. Brook$^{1,2}$ \thanks{E-mail: \href{mailto:}{chbrook@ull.edu.es}}, M. Huertas-Company$^{2,1,3}$, A. Di Cintio$^{1,2}$,\newauthor
       A.V. Macci\`{o}$^{4,5,6}$, R. J. J. Grand$^{2,1}$ and G. Battaglia$^{2,1}$\\
$^{1}$Universidad de La Laguna. Avda. Astrof\'{i}sico Fco. S\'{a}nchez, La Laguna, Tenerife, Spain\\
$^{2}$Instituto de Astrof\'{i}sica de Canarias, Calle Via L\'{a}ctea s/n, E-38206 La Laguna, Tenerife, Spain\\
$^{3}$ LERMA, Observatoire de Paris, CNRS, PSL, Universit\'{e} de Paris, France\\
$^{4}$ New York University Abu Dhabi, PO Box 129188 Abu Dhabi, United Arab Emirates \\
$^{5}$ Center for Astro, Particle and Planetary Physics (CAP$^3$), New York University Abu Dhabi \\
$^{6}$ Max Planck Institute f\"{u}r Astronomie, K\"{o}nigstuhl 17, D-69117 Heidelberg, Germany 
 \\
}

\begin{document}

\date{Accepted xxxx. Received xxxx; in original form xxxx}

\pagerange{\pageref{firstpage}--\pageref{lastpage}} \pubyear{2020}

\maketitle

\label{firstpage}

\begin{abstract}
Numerical simulations within a cold dark matter (DM) cosmology form halos whose   density profiles have a steep  inner slope  (`cusp'),  yet observations of galaxies often point towards a flat central `core'.
We develop a convolutional mixture density neural network model  to derive a probability density function (PDF) of the inner density slopes of DM halos. We train the network on simulated dwarf galaxies from the NIHAO and AURIGA projects, which include both DM cusps and cores:  line-of-sight velocities and  2D spatial distributions of their stars are used as inputs to obtain a PDF representing the probability of  predicting a specific inner slope.
The model recovers accurately the expected DM profiles: $\sim$82$\%$ of the galaxies have a derived inner slope within $\pm$0.1 of their true value, while $\sim$98$\%$  within $\pm$0.3. We apply our model to four Local Group dwarf spheroidal galaxies and find results consistent with those obtained with the Jeans modelling based code {\sc GravSphere}: the Fornax dSph has a strong indication of possessing a central DM core, Carina and Sextans have cusps (although the latter with  large uncertainties), while Sculptor shows a double peaked PDF indicating that a cusp is preferred, but a core can not be ruled out.  Our results show that simulation-based inference with neural networks provide a innovative and complementary method for the determination of the inner matter density profiles in galaxies, which in turn can  help constrain the properties of the elusive DM.
\end{abstract}

\noindent
\begin{keywords}
galaxies: dwarf - evolution - formation - halos - dark matter
\end{keywords}

\section{Introduction} \label{sec:introduction}
Dark matter (DM) halos that form in simulations within a $\Lambda \rm CDM $ cosmological context have a characteristic density profile, which has a logarithmic inner slope of -1  \cite[the NFW profile]{navarro96}. Such a steep inner density profile has been referred to as a `cusp'. Nevertheless, observations of dwarf galaxies inhabiting these halos have shown discrepancies with the predictions of the model, showing significant evidence that several of these galaxies have a flat inner density profile, with slope approaching zero, referred to as a `cored' profile \citep{moore94}.
The discrepancy between theory and observations has been referred to as the `core-cusp' problem \cite[e.g.][]{simon05,deblok08,bullock17}. 

While over the years several alternative DM models have been proposed to tackle this issue \cite[e.g.][]{spergel00,kaplinghat16,scheider17}, 
it has  been also shown that cores can be explained  within  $\Lambda \rm CDM $  considering the effect that baryons have on DM matter. \cite{navarro96b} showed that if gas is slowly accreted onto a dwarf galaxy and then suddenly removed through processes such as stellar winds or supernovae feedback, the DM distribution can expand, lowering the  central density of the halo. This effect of DM heating is small in realistic conditions \citep{gnedin02}, but \cite{read06} showed that if the effect repeats over several cycles of star formation, it accumulates leading to a complete core formation. This core can be permanent if the outflows are sufficiently rapid \citep{pontzen12}. Modern hydrodynamical simulations of dwarf galaxies that take into consideration baryonic feedback and have a sufficiently high  density threshold for star formation have indeed succeeded at creating DM cores  \cite[e.g.][]{governato10, DiCintio2014a, tollet15,chan15}.
Still, the ‘cusp-core’ problem is far from being completely
solved, due to the difficulties of uncovering the underlying
DM distribution in observed dwarf galaxies, and significant effort has gone into the development and improvement of methods to infer the inner DM density profile of such galaxies.

Analysis of the rotation velocity of gas in low surface brightness galaxies, for example,  allow to derive and fit their underlying DM distribution 
suggesting the presence of a DM core in such systems \citep[e.g.][]{moore94,gentile04,deblok08,lelli16}.
On the other side, in pressure-supported galaxies which are devoid of gas,
 such as the dwarf spheroidal galaxies (dSphs) found within the Local Group,
 the  kinematic information on which  dynamical modeling   relies on, comes from the line-of-sight velocity distribution of their stellar component. 

 A variety of methods have been employed on dwarf galaxies to derive their central DM density, such as Jeans \citep[e.g.][]{Marel94,kleyna01,battaglia08,ReadJustin19,Collins21} or Schwarzschild \citep[e.g.][]{Schwarzschild79,cappellari06,bosch10,Breddels13a,Breddels13b} modelling.
The results in the literature seem to point to cored DM profiles being favoured over cuspy ones in the Fornax dSph \citep[e.g.][]{Geha06,Walker11,brook15a,Pascale18}, while in the case of Sculptor, another very well studied system, it is  still very much debated if its DM halo is cored or cuspy, perhaps pointing to the presence of a mild cusp \citep[e.g.][]{Breddels13b,Zhu16,Hayashi20} (for a review on these topics,  see \citealt{Battaglia22} and references therein).

A central limitation of  the previously mentioned models, however, comes from the uncertainty in the anisotropy of the stellar orbits, in the case of  Jeans modelling, which causes a degeneracy with the underlying mass profile \citep{Binney82};
Schwarzschild modelling, on the other end, is hampered by its sensitivity
to the available data \citep{Kowalczyk17}.

In this work we present an alternative and innovative method to discriminate between cusps and cores in dwarf galaxies based on machine learning techniques. Namely, we use convolutional mixture density neural networks to determine a posterior distribution of the inner profile of DM halos. This general approach has been successfully implemented for measuring cluster masses from galaxy dynamics \cite[e.g.][]{ho19,ramanah20a,ramanah20b}.
The neural network uses as inputs the phase-space mappings of positional and dynamical distributions of stars within galaxies. 

We use a suite of 171 dwarf galaxies from the NIHAO project with different initial conditions and parameters \citep{wang15,dutton20} and 12 dwarf galaxies from the AURIGA project \citep{Grand17} as a training set for the network. 
We then apply our novel model to four dwarf spheroidal galaxies satellites of the Milky Way to infer the inner slope of their DM density profiles.

The paper is organised as follows. In section 2 we present the simulation data set and the machine learning architecture. 
In section 3 we show the results of the trained  model on the test set, and  we then apply the model to observed dwarf galaxies in section 4. The conclusions  are discussed in section 5.

\section{Methods} \label{sec:methods}

\subsection{The Training Set}\label{sec:simu}
To train our model we need a large set of simulated dwarf galaxies with well known density profiles. We use fully cosmological simulations from NIHAO \citep{wang15} and AURIGA \citep{Grand17} projects, in which  DM  and baryonic matter evolve together, making our training set as realistic as possible. 

\begin{figure}  
\includegraphics[width=\columnwidth]{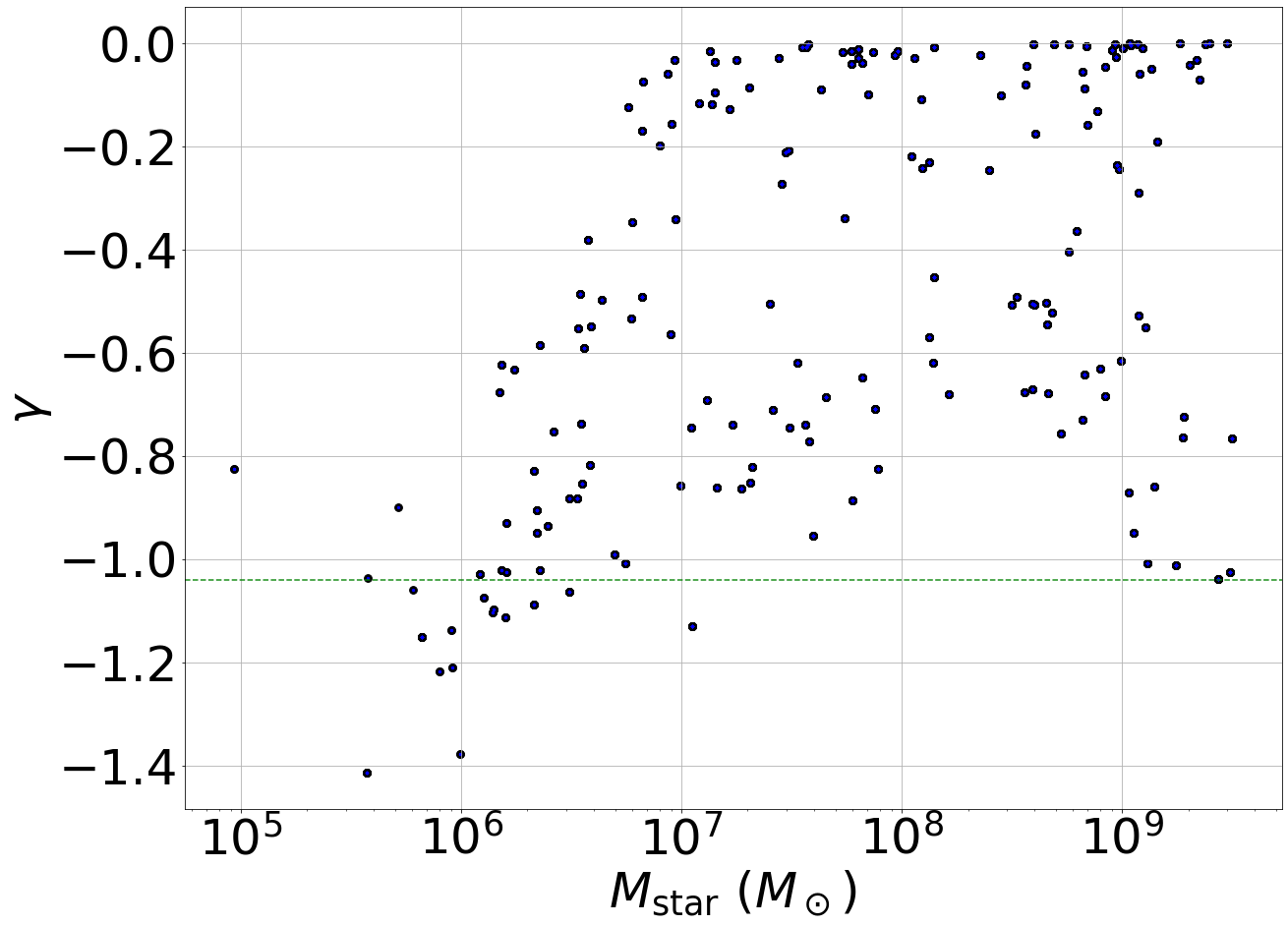}
\caption{Relationship between the inner slope of the DM density profiles $\gamma$ (defined as the logarithmic slope at 150 pc) and  the stellar mass of the simulated galaxies in our dataset. The green horizontal line marks the value of $\gamma$ for a NWF profile.}
\label{fig:gamma-M} 
\end{figure}
Importantly, we need to include simulations of galaxies with  both cusps and cores in their central region, and with various stellar masses,  in order to minimise any systematic dependence of cusp and core on    properties such as mass. Indeed, the fiducial NIHAO galaxies have a density profile highly correlated with mass \citep{DiCintio2014b,maccio20}, which would allow the machine learning code to predict cusp or core based on any indicator of total mass, rather than by the details of the stellar velocities and positions. We therefore use simulations that have a range of different physical and/or parametric inputs, meaning that our final suite of simulations includes a range of inner slopes at various masses and sizes. 
We firstly include dwarf galaxies within the fiducial NIHAO model, ranging in halo mass from $\sim$  $10^9$ \msun \ to $10^{11.5}$ \msun \ and stellar mass from an order of $10^5$ \msun \ to $10^{9.5}$ \msun. This model includes energy feedback from massive stars and supernovae \citep{stinson06}, which has been shown to be able to modify the inner density profile and result in cores, particularly in simulated galaxies with stellar mass between $10^{7}$\msun and $10^{9}$\msun \ \citep{DiCintio2014a}.
We also use  simulations of dwarfs from \cite{dutton20}, that employs the same model as the fiducial NIHAO ones, but with different star formation thresholds, ranging from $\rho_{\text{thresh}}$ = 0.1 to 100 particles per cm$^{-3}$: this translates into galaxies of a similar stellar mass  ending up with different density profiles, as the star formation density threshold has been shown to be one of the most important parameter for core formation in baryonic simulations (see \citealt{Benitez19, dutton20}). 
We further add a set of simulations with no stellar feedback run from the same initial conditions as fiducial NIHAO \citep{wang15}. The lower total feedback energy results in different inner density profiles than simulations in which the stellar feedback is included, for the same initial conditions, therefore further increasing the desired diversity of central DM profiles at a given galaxy mass.
Finally, we include 12 simulated dwarf galaxies from the AURIGA project \citep{Grand17}, all of which have a central DM cusp.

We have 183 simulated dwarf galaxies in total: 60 simulations from the fiducial NIHAO suite \citep{wang15}, 101 simulations from \cite{dutton20} with varying density thresholds and varying density profile, 10 simulations without stellar feedback also from \cite{wang15} and 12 simulations from \cite{Grand17}. All together, these simulations have a range in halo mass between \mhalo=$3\cdot10^9$ \msun and \mhalo=$4\cdot10^{11}$ \msun.  NIHAO simulations resolve the mass profile of galaxies to below 1 per cent of their virial radius at all masses, while AURIGA simulations are constructed to have a maximum physical softening of $\sim 370$ pc.

We define the  DM inner slope value of the simulated galaxies as the slope at 150 pc of the DM density profile of each galaxy. This value is extrapolated from the fit of the density profile to a double-power law profile \citep{DiCintio2014b}, in order to avoid the noise effect of the computed density profile of the simulations in inner regions very close to the softening length \footnote{This extrapolation is reasonable considering that  AURIGA galaxies, although not resolved at r$<$370 pc, consistently show a cuspy inner density, i.e. there is no sign of an artificial central DM core.}.
We end up with a set of simulated dwarfs exhibiting a range of density profiles: the relationship between stellar mass and inner slope of DM halo for our full dataset can be seen in Fig. \ref{fig:gamma-M}.

\begin{figure}
\includegraphics[width=\columnwidth]{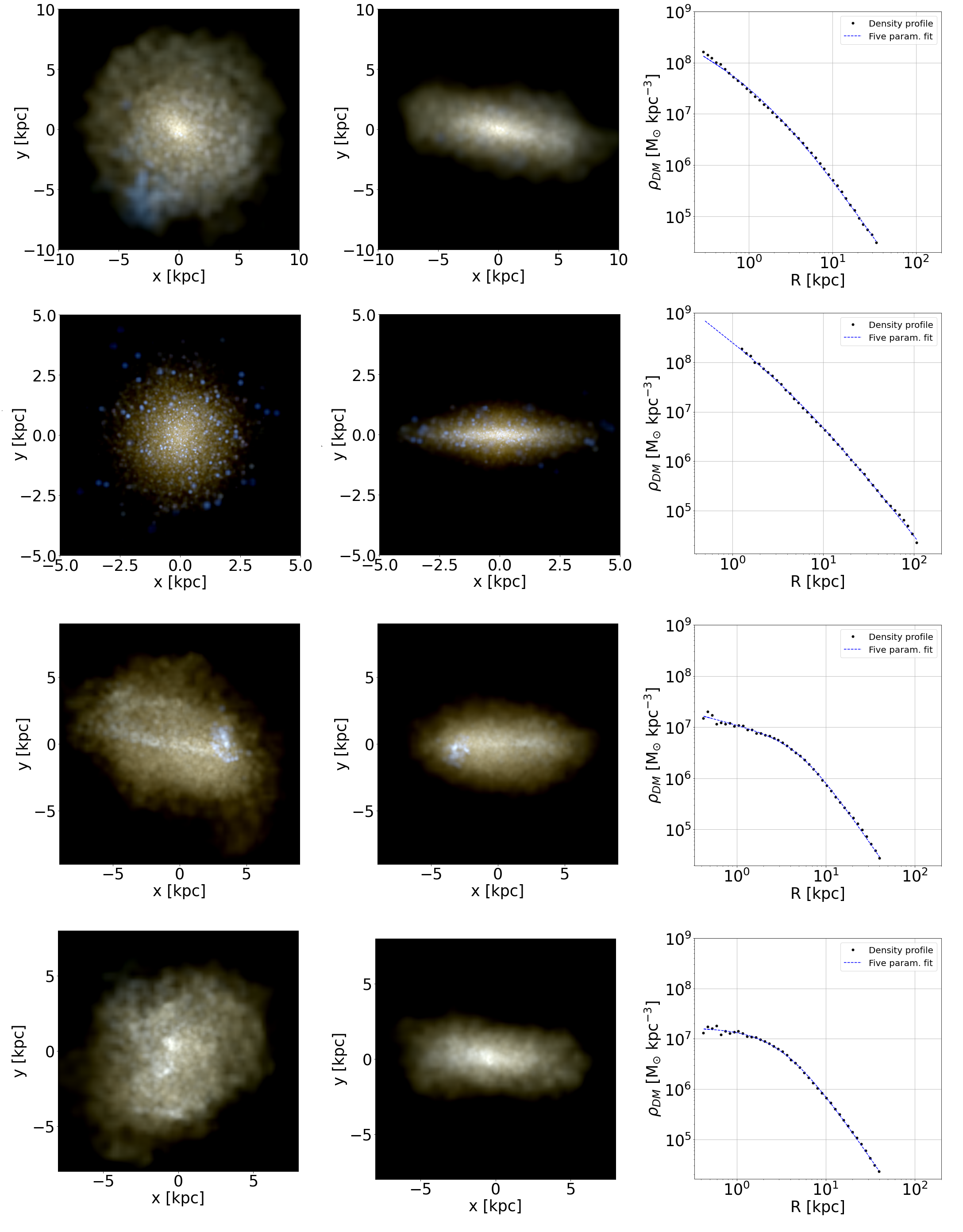}
\caption{Example of cored and cuspy galaxies from our simulation dataset. Here, each row represents a different galaxy. Left columns: rendering of the stars in a face-on view. Central columns:  rendering of the stars with a edge-on orientation. Right columns: DM density profiles and fit to a double-power law model \citep{Jaffe83,Merritt06}.}
\label{fig:render_gal} 
\end{figure}
To increase the size of our training set, we use 3 different output timesteps for each galaxy: z=0, z=0.112, and z=0.226. Each  simulated galaxy is already virialized at  these redshifts, and it is therefore possible to take  different snapshots of the dwarf. While this procedure does not change greatly the range of the obtained density profiles, it does change the position and velocities of the stars within each galaxy. We end up with a sample of 549 galaxy snapshots which we will use as training set for our method. We show in Fig. \ref{fig:render_gal} examples of stellar renderings of cored and cuspy simulated galaxies together with their corresponding DM profiles. 
We then proceed to select stars within each galaxy snapshot.
Typically, the number of stars for which spectroscopic data is available for Local Group dwarf galaxies is the order of hundreds or thousands, while the number of star particles available in our simulated galaxies range  from a few hundred to several million, with a mean number of about  $10^5$ stellar particles in each galaxy. 

Therefore, in order to simulate an observational sample of stars,  and to further expand our training set, we have divided each simulated galaxy's complete sample of stars into a minimum of 20 subsets, each made of randomly selected stars. The number of stars within each subset of a given galaxy is dependent on the total number of star particles in the simulation,  with  an upper limit of $10^4$ stars and a lower limit of 200 stars.
The stars of each subset are then projected in arbitrary sky planes to simulate galaxies observed from different viewing angles. These projected stars are defined by their position ($x_{\rm proj}$, $y_{\rm proj}$) and their line-of-sight velocity $v_{\rm LOS}$. We oversample some galaxies by making multiple projections to each of their subset, and undersample some galaxies, with the objective of making the training set have a uniform distribution of inner slopes: this avoids biases in the model during training. 
We end up with a  total of 10273  data sets to train our model, each composed of randomly selected stars within different simulated galaxies and at different viewing angles, for which we stored information about their positions ($x_{\rm proj}$, $y_{\rm proj}$) and line-of-sight velocities $v_{\rm LOS}$.

\subsection{The information inputs}

The inputs of our deep neural network model are continuous 2D probability density functions (PDFs) of the distribution of stars in projected phase spaces, constructed with bivariate kernel density estimations (KDEs). The mapping generated with KDEs allows us to encapsulate the features of the original discrete distributions in the same form even if each galaxy subset is represented by a different number of stars.

\subsubsection{Kernel Density Estimation}
\label{sec:inputs}
Let $\mathbf{X_1}$, $\mathbf{X_2}$, ..., $\mathbf{X_n}$ denote a sample of size $n$ from a random variable with density $f$, each variable being a two-dimensional vector for the case of a bivariate KDE. The kernel density estimate of $f$ at the point $\mathbf{x}$ is given by

\begin{equation}
    f_h(\mathbf{x}) = \frac{1}{n|\mathbf{H}|^{1/2}}\sum^n_{i=1}K\left[\mathbf{H}^{-1/2}(\mathbf{x-X_i})\right],
\end{equation}

\noindent where K is a kernel function and $\mathbf{H}$ is a 2x2 bandwidth matrix. 

\begin{figure*}
\includegraphics[width=6.81in]{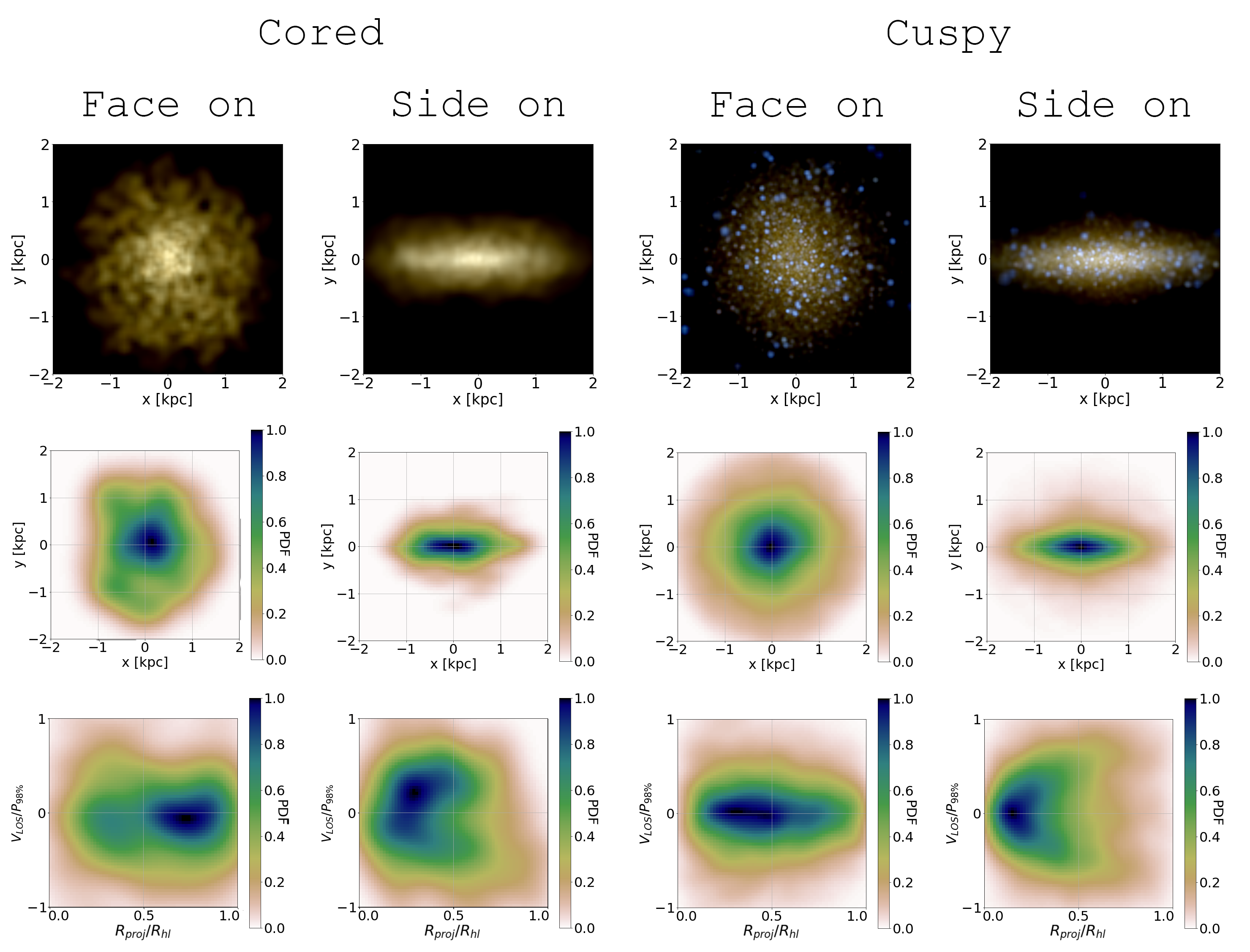}
\caption{Model inputs for a cored and a cuspy galaxy, each one represented face-on  and edge-on. The logarithmic slope at 150 pc is $\gamma = -0.20$ for the cored galaxy and $\gamma = -1.32$ for the cuspy galaxy. From top to bottom: A 3-color image of the stars in the galaxy; the PDF in the \{x,y\} phase space; and the PDF in the \{$\hat{R}_{\rm proj},\hat{v}_{\rm LOS}$\} phase space.}
\label{fig:example_inputs} 
\end{figure*}

The KDE sums up the density contributions from the collection of data points at the evaluation point $\mathbf{x}$, so that data points close to $\mathbf{x}$ contribute significantly to the total density, while data points further away from $\mathbf{x}$ contribute less. The shape of those contributions is determined by K, and their dimensions and orientation by $\mathbf{H}$.
Usually the kernel function K is chosen to be a probability density symmetric about zero \citep{Sheather04}. In this work we use a 2D Gaussian kernel:

\begin{equation}
    K(\mathbf{u}) = (2\pi)^{-3/2}|\mathbf{H}|^{1/2} exp\left(-\frac{1}{2}\mathbf{u}^T \mathbf{H}^{-1}\mathbf{u}\right),
\end{equation}

\noindent where $\mathbf{u} = \mathbf{x - X_i}$. For the bandwidth matrix, a scaling factor $\kappa = n^\frac{-1}{6}$ is multiplied by the covariance matrix of the data, where $n$ is the number of data points.

\subsubsection{Model inputs}
\label{sec:pdfs}

From the projected information (positions in the x-y plane and $v_{LOS}$) of the sample of stars representing each galaxy we have made two maps:

\begin{itemize}
    \item A PDF sampled at 64x64 points with the distribution of stars in \{x,y\} phase space, between -2 kpc and 2 kpc in each coordinate, in the reference system where (x,y) = (0,0) is the center of the galaxy.
    \item A PDF sampled at 64x64 points with the distribution of stars in  \{$\hat{R}_{\rm proj},\hat{v}_{\rm LOS}$\} phase space, where $\hat{R}_{\rm proj} = \sqrt{x^2 + y^2}/R_{\text{hlr}}$ is the radial position normalized by the half-light radius $R_{\text{hlr}}$ and $\hat{v}_{\rm LOS} = v_{\rm LOS}/P_{98\%}$ is the line-of-sight velocity normalized by the 98\% percentile of the absolute value of $v_{\rm LOS}$ of all  stars of the sample. $\hat{R}_{\rm proj}$ ranges  from 0 to 1, and $\hat{v}_{\rm LOS}$ ranges from -1 to 1.
\end{itemize}

In Fig. \ref{fig:example_inputs} we show our model inputs, as the PDFs corresponding to both maps, for a cored (left) and cuspy (right) galaxy.

\begin{figure*}
\includegraphics[width=6.5in]{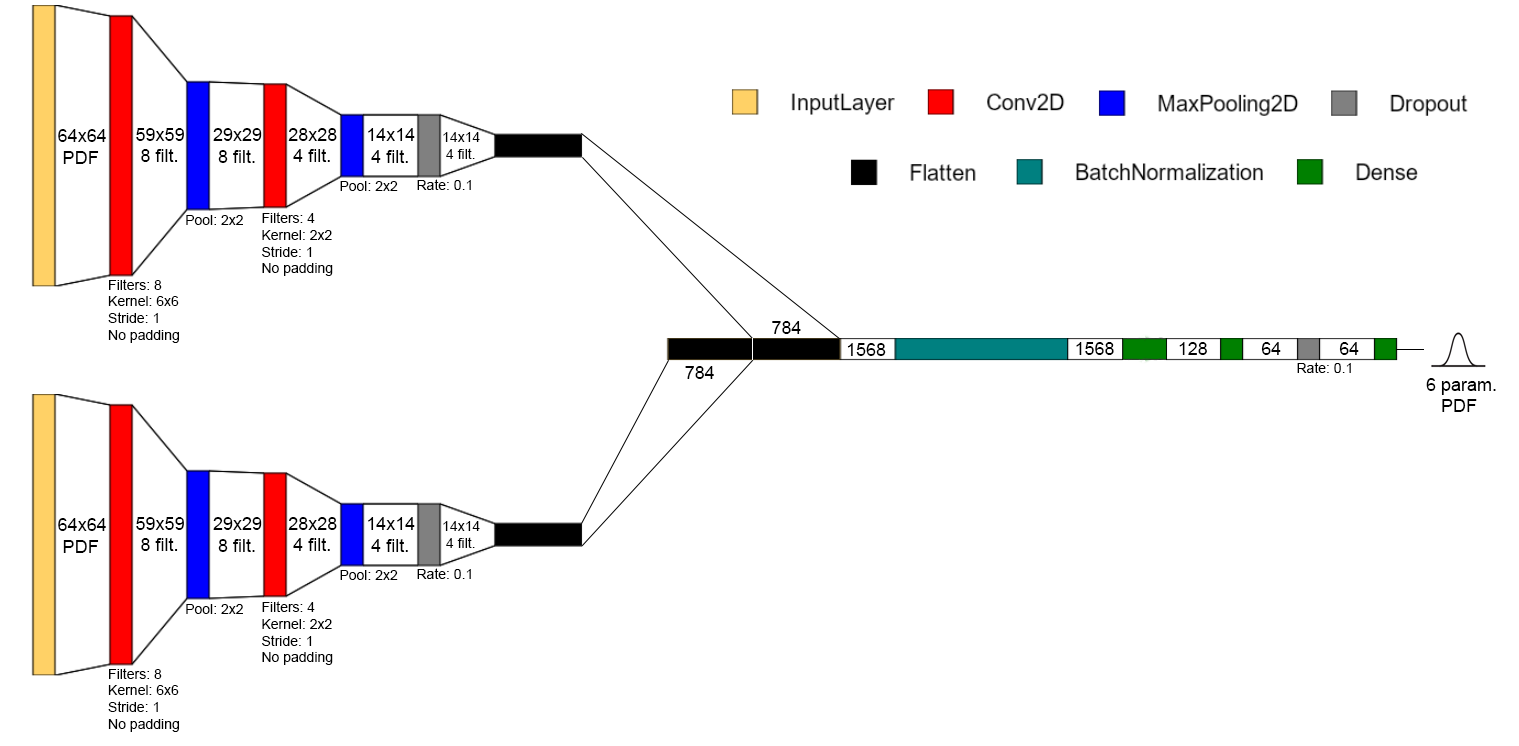}
\caption{Schematic representation of our double channel MDCNN architecture to infer inner slopes of the DM profiles (slope at 150 pc) of galaxies from their 2D phase-space mappings of positional and dynamical distributions of stars. The MDCNN extracts the spatial features from the phase-space mappings and gradually compresses into high-order features until describing the input with only 6 parameters, which are used as parameters of a double Gaussian corresponding to the probability density distribution of the inner slopes values.}
\label{fig:architecture} 
\end{figure*} 

\subsection{The model}

In this work, we use mixture density convolutional neural networks (MDCNNs) to map the input data composed of the two PDFs described in Section \ref{sec:inputs} into the inner slopes of the DM profiles of the galaxy associated to those two PDFs. We approximate the posterior distribution of the slopes with the sum of two dimensional Gaussian distribution whose parameters are estimated by the neural network. \footnote{The use of a double Gaussian yields more accurate predictions than using a single one. On the other hand, using more than two Gaussians does not lead to more accurate slope predictions.} Our model takes as input a two channel image consisting of  the PDFs on the \{$\hat{R}_{\rm proj},\hat{v}_{\rm LOS}$\} phase space and the \{x,y\} phase space separately. The images are passed through 2 convolutional sequential layers. The outputs of the two convolutional branches are then concatenated and fed into a 3 layer fully connected network. The final output consists of 6 parameters which parametrize the joint 2D Gaussian posterior.

A schematic view of the architecture used in this work can be seen in Fig. \ref{fig:architecture}, while a more in-depth description of the different layers and neural network methods can be found in Appendix A.

\subsubsection{Training and evaluation}

The training is done over a training set consisting of 10273 galaxy subsets with their respective inner slopes, which act as targets. The loss function to minimize during the training is the negative logarithmic likelihood of the training sample, defined as

\begin{equation}
    L = -\ln L* = -\sum^N_{i=1} \ln\left[p_{g}(t_i|\theta)\right],
\end{equation}

\noindent where $t_i$ is the inner slope of the galaxy subset $i$ and $\theta$ the set of parameters of the distribution $p_g$. For a certain galaxy subset, the likelihood is the value of the probability density function (defined with a multivariate Gaussian distribution as the output of the last layer) in its real inner slope value; i.e. the probability the model predicts for the inner slope of the galaxy to be its correct value:

\begin{equation}
    p_{g}(x|\theta) = \sum^{2}_{j=1} \phi_j \mathbf{N}(x, \mu_j, \sigma_j),
\end{equation}

\noindent where $\mathbf{N}(x, \mu_j, \sigma_j)$ is the $j$ Gaussian with mean $\mu_j$ and standard deviation $\sigma_j$, $\phi_j$ is the weight of the $j$ Gaussian, so that $\sum^{n_g}_{j=1} \phi_j = 1$, and $\theta$ is then a set of six parameters (mean, standard deviation and weight of the two Gaussians).

The minimization of the loss function is done with the adaptative moment estimation ({\small ADAM}) optimizer, an algorithm for optimization that uses the gradient descent iterative technique. Between the popular learning-method algorithms, {\small ADAM} is shown to compare favorably in performance and computational cost \citep{Diederik14}.
After training, the evaluation of the model outputs a multivariate Gaussian distribution that can be understood as an approximation to the true posterior distribution of the inner slope of a given input,  given the prior distribution of the inner slopes in the training dataset. This posterior then represents the probability that the model assigns  a certain value of the inner slope given the set of observables under the prior of the training set.

Usually, the test dataset for the final evaluation of the converged model is constructed by randomly taking a sufficient number of elements from the complete dataset to correctly represent all feature variety in the data. In this work, due to the limited number of galaxies available, removing too many galaxies with varying characteristics from the training dataset is expected to worsen the performance of the model, since we do not have many different examples of galaxies with similar characteristics to each other. To properly evaluate the model, we have performed multiple complete training runs using only 10 galaxies as validation and test datasets in each one, changing the galaxies that would come out of the training dataset in each of the training runs to evaluate the network in several projections of every galaxy. This allows us to analyse the consistency of the model training and its performance in a large number of galaxies without compromising the training dataset.

\subsubsection{Representing uncertainties}

The output posterior distribution represents the random or aleatoric  uncertainty in the slope prediction of the final model, but it does not represent the uncertainty due to the stochastic nature of the weight determination while training the neural network (epistemic uncertainty), which can lead to different models for the same training conditions when dealing with limited data. 
We use the Monte Carlo dropout method (MC-Dropout) \citep{Gal15} to approximate the epistemic uncertainty which is based on the repeated evaluation of the same input, randomly setting to 0 the weights on some layers while doing each inference, to construct a final evaluation with statistical information about the epistemic uncertainty. \cite{Gal15} showed that applying dropout during inference is equivalent to an approximation to a probabilistic Deep Gaussian process. It means we can measure the epistemic uncertainty by applying the dropout layer during inference for a statistically relevant number of them, acquiring a predictive mean and variance for each point of the posterior distribution. The constructed final posterior for each galaxy projection is the normalized mean of 100 multivariate Gaussian posteriors inferred by the model with active dropout layers.

\begin{figure}
\includegraphics[width=\columnwidth]{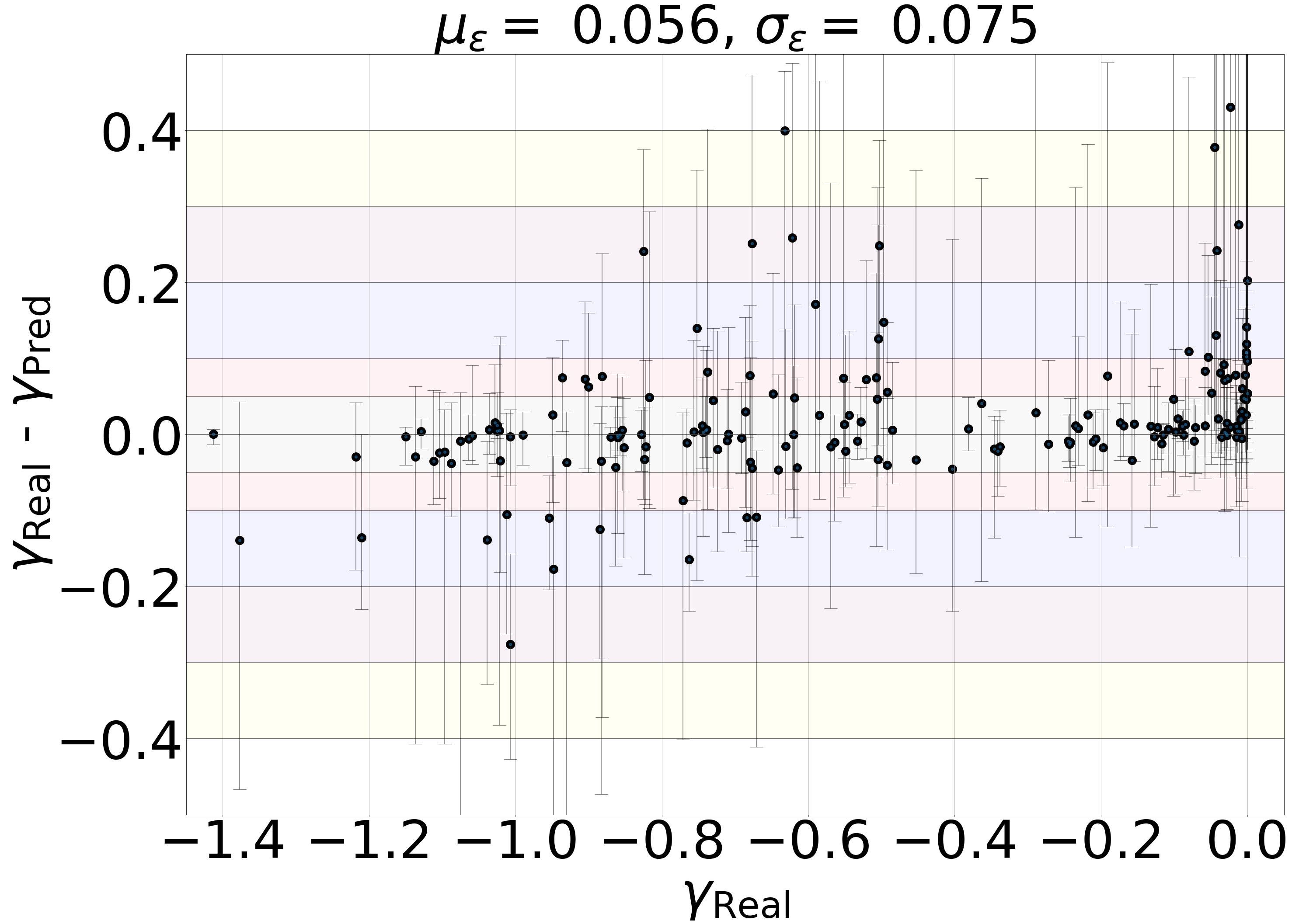} \\
\caption{Difference between real and predicted value of DM profiles inner slopes (defined at 150 pc) vs real inner slope, for the simulated galaxies used in this work, defining the predicted value as the mode of the posterior distribution. Each point represents the mean $\gamma_{\rm Real} - \gamma_{\rm Pred}$  for all the projections of each individual galaxy, while error bars span the range between the minimum and the maximum deviation amongst every possible projection of each galaxy. Colored areas represent increasing deviation ranges, from 0.05 to 0.4.}
\label{fig:acc_error} 
\end{figure}
\vspace{-.2cm}
\section{Results} \label{sec:results}
The goal of our work is to infer the logarithmic inner slope of the mass density profile in the central region of a galaxy (from now on: inner slope) from spectroscopic data of a random sample of its stars. To do so, all simulated galaxies and their subsets of stars are randomly projected in several sky planes, to simulate several viewing angles, and the neural network is trained to infer the inner slope of the galaxy from the positions  and line-of-sight velocities of its stars. For each galaxy the neural network outputs a probability density function which approximates the posterior probability of obtaining a specific inner slope given the inputs.

\subsection{Predicting DM inner slopes}

We define two different methods to construct the predicted slope value $\gamma$ from the posteriors:

\begin{itemize}
    \item by using the mode of the posterior distribution (i.e. the maximum of the PDF): $\gamma_{\rm Pred,mode}$.
    \item by using the mean of the normalized posterior distribution: $\gamma_{\rm Pred,mean}$.
\end{itemize}

The deviation $\epsilon$ of a prediction from its true value is defined as $\epsilon_i = \gamma_{\rm Real} - \gamma_{\rm Pred,i}$, where $\gamma_{\rm Real}$ is the real slope at 150 pc of the DM profile of a galaxy simulation.
The  results for the mode method can be seen in Fig. \ref{fig:acc_error}, which shows the difference between the real  and predicted slopes of our simulated dwarf galaxies, $\gamma_{\rm Real} - \gamma_{\rm Pred}$, as a function of the real slope.
Each point represents the mean deviation for every projection of each individual galaxy, while the deviation bars indicate the minimum and the maximum value amongst every possible projection of each galaxy. Shaded colored horizontal areas represent increasing uncertainty ranges, from $\pm$0.05 to $\pm$0.4.

The mean global absolute deviation on the predicted inner slope, for all the galaxies in our set, is of $\mu_{\epsilon} = 0.056$ for the mode method and of $\mu_{\epsilon} = 0.068$ for the second method.
Note that while cuspy and `in between' galaxies are scattered around $\gamma_{\rm Real} - \gamma_{\rm Pred}$ = 0, cored galaxies tending towards $\gamma=0$ are necessarily only scattered at  $\gamma_{\rm Real}- \gamma_{\rm Pred} \geq $ 0, since by construction the maximum possible inner slope is 0.

\begin{table}
\begin{center}
\begin{tabular}{ccc}
\hline
\hline
Deviation range ($\pm$ X) & \% of projections with  & \% of galaxies with\\
 & $|\epsilon| \leq X$ &$|\epsilon| \leq X$\\
\hline
0.05 & 66.67 & 67.80\\                   
0.1 & 80.79 & 81.92\\ 
0.2 & 94.35 & 94.35\\                   
0.3 & 98.31 & 98.31\\ 
0.4 & 99.44 & 98.87\\ 
\hline
\hline
\end{tabular} \\
\caption{Percentage of all the projections (central column) and of individual galaxies (right column) whose  predicted inner slope lies  within a given deviation range X, i.e.  $|\epsilon| = |\gamma_{\rm Real}- \gamma_{\rm Pred,mode}| \leq X$. Here, we used the mode of the posteriors method to derive the inner slopes.}
\label{tab:ac_error}
\end{center}
\end{table}

\begin{figure}
\includegraphics[width=\columnwidth]{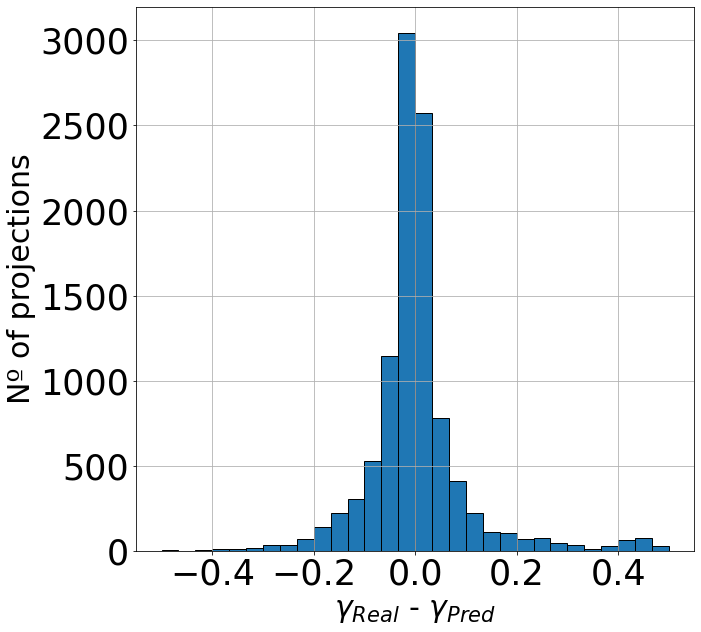} \\
\caption{Distribution of $\gamma_{\rm Real}- \gamma_{\rm Pred,mode}$ for every projection of each galaxy in our training set.}
\label{fig:hist_error} 
\end{figure}
In Table \ref{tab:ac_error} we can see the percentages of correctly predicted inner slopes, taking into account all the projections of every galaxy (middle column) and each galaxy individually (right column), for our  complete test dataset, within several uncertainty ranges.
Roughly 82$\%$ of the galaxies recover the correct, real inner slope within $\pm0.1$, while  98\% of them lie within $|\gamma_{\rm Real} - \gamma_{\rm Pred}|\leq 0.3$.
These ranges are clearly small enough to shed light on the discussion regarding the presence or not of cores in dwarf galaxies.
 
Finally, a histogram of the deviation distribution for every projection of each galaxy (i.e. 10273 in total) can be seen in Fig. \ref{fig:hist_error}, indicating that the values of $\gamma_{\rm Real} - \gamma_{\rm Pred}$ are peaked at and roughly symmetrically distributed around 0, except for very cored galaxies that have by definition $\gamma_{\rm Real} - \gamma_{\rm Pred} \geq 0$, as already stated, and a small asymmetry towards predicting stronger cores in galaxies in the range of small deviations.
We  showed that our method predicts accurately the expected inner slope of galaxies regardless of their actual real slope, with a mostly uniform scatter of $\sigma_\epsilon = 0.075$.

\begin{figure}
\includegraphics[width=\columnwidth]{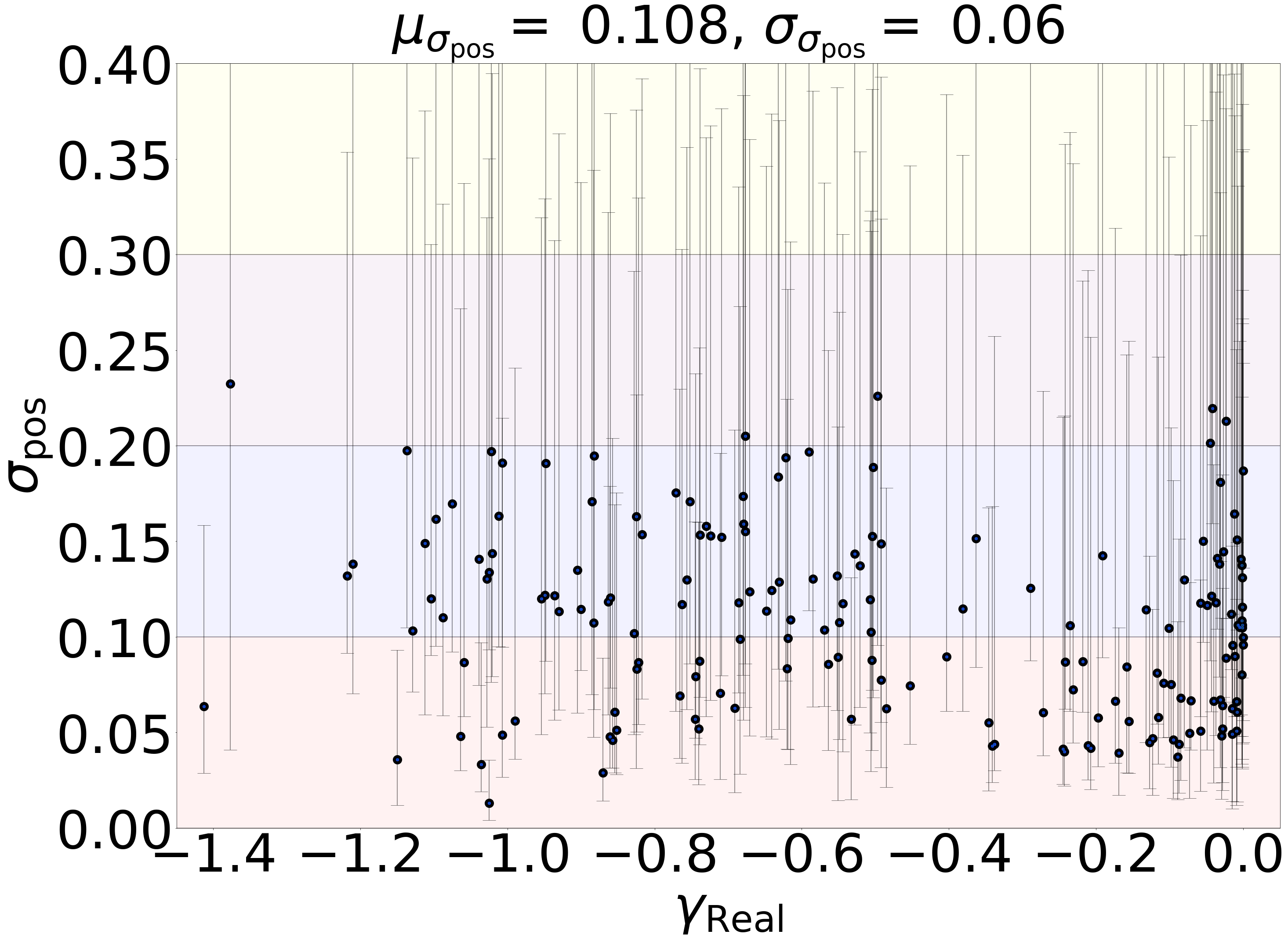}
\caption{Standard deviation $\sigma_{\rm pos}$ of each posterior PDF vs $\gamma_{\rm Real}$, for the simulated galaxies used in this work. Each point represents the mean standard deviation of each posterior, for every  projection of an individual galaxy. The error bars range between the minimum and maximum standard deviation value of the posteriors of all the projections of that galaxy.}
\label{fig:sigmas} 
\end{figure}

\subsection{Uncertainty in the inference}
In Fig. \ref{fig:sigmas} we show the standard deviation $\sigma_{\rm pos}$ of each posterior PDF from every  galaxy in the test dataset, defined as the square root of the variance of the normalized posterior:

\begin{equation}
    \sigma^2_{\rm pos} = \int_{-\infty}^\infty (\gamma - \mu_{\rm pos})^2 P(\gamma) \ d\gamma,
\end{equation}

\noindent where $P(\gamma)$ is the normalized posterior distribution and $\mu_{\rm pos}$ is the mean of the distribution:

\begin{equation}
    \mu_{\rm pos} = \int_{-\infty}^\infty \gamma P(\gamma) \ d\gamma.
\end{equation}

\begin{table}
\begin{center}
\begin{tabular}{cc}
\hline
\hline
Region & \% of projections for which $\gamma_{\rm Real}$\\
& is within region\\

\hline
$1-\sigma_{\rm pos}$ & 86.29 \\                   
$2-\sigma_{\rm pos}$ & 97.57 \\ 
$3-\sigma_{\rm pos}$ & 99.73 \\                   
\hline
\hline
\end{tabular} \\
\caption{Percentage of predictions within increasing $\sigma_{\rm pos}$ ranges X, defined as $|\gamma_{\rm Real}- \gamma_{\rm Pred}| \leq X$.}
\label{tab:coverage}
\end{center}
\end{table}
\noindent The mean of all the $\sigma_{\rm pos}$ of the dataset, $\mu_{\sigma_{\rm pos}}$, is around 0.1 and only 8.99\% of the projections have values of $\sigma_{\rm pos}$ greater than 0.2, uncertainties that are  small enough to clearly distinguish between cores and cusps in the vast majority of cases.
Fig. \ref{fig:sigmas} shows that the standard deviation $\sigma_{\rm pos}$ of each posterior PDF is uniform across the inner slopes values, i.e. the  width of the PDFs  does not depend on the inner slope of  galaxies, such that the model is not biased towards recovering with higher accuracy either cusps or cores.
Most galaxies show a significant variation in the size of their uncertainties depending on the projection, indicating that the amplitude of the uncertainty is strongly correlated with the angle of observation.

\begin{figure}
\includegraphics[width=\columnwidth]{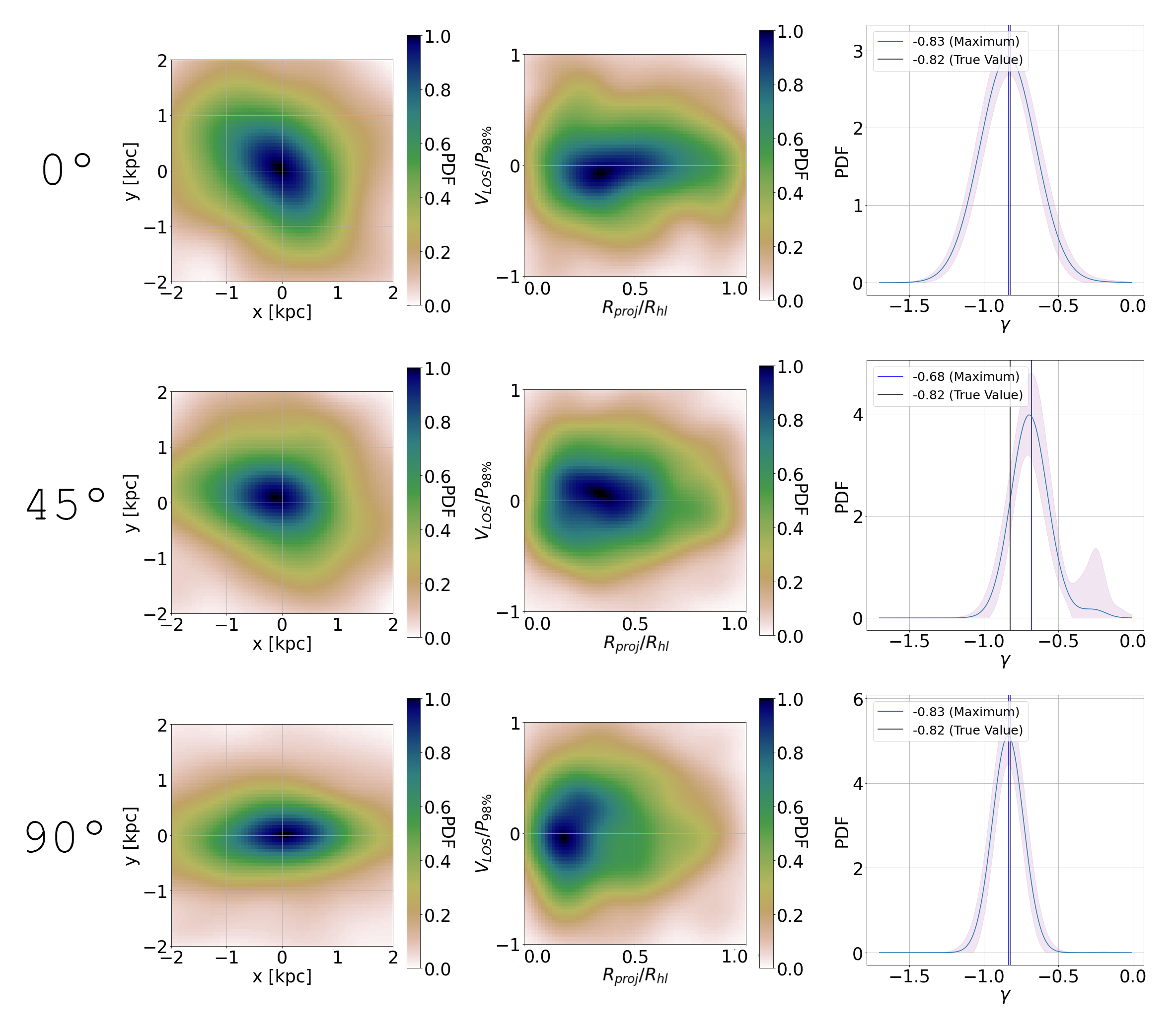}
\caption{Probability density distributions used by the neural network as  input in the case of one simulated galaxy subset seen at 0º (face-on), 45º and 90º (side-on), alongside with the Bayesian posteriors predicted by the model. Left columns: PDFs in the \{x,y\} phase space. Central columns: PDFs in the \{$\hat{R}_{\rm proj},\hat{v}_{\rm LOS}$\} phase space. Right columns: predicted Bayesian posterior in the space of inner slope of the DM profile (slope at 150 pc); shaded regions represent the standard deviation of the posterior values for the MC-Dropout inferences at each slope point, while  the blue vertical line shows the mode (maximum) of the posterior distribution and the black one the true value of the inner DM slope.}
\label{fig:example_posteriors_1} 
\end{figure} 

\begin{figure*}
\includegraphics[width=6.95in]{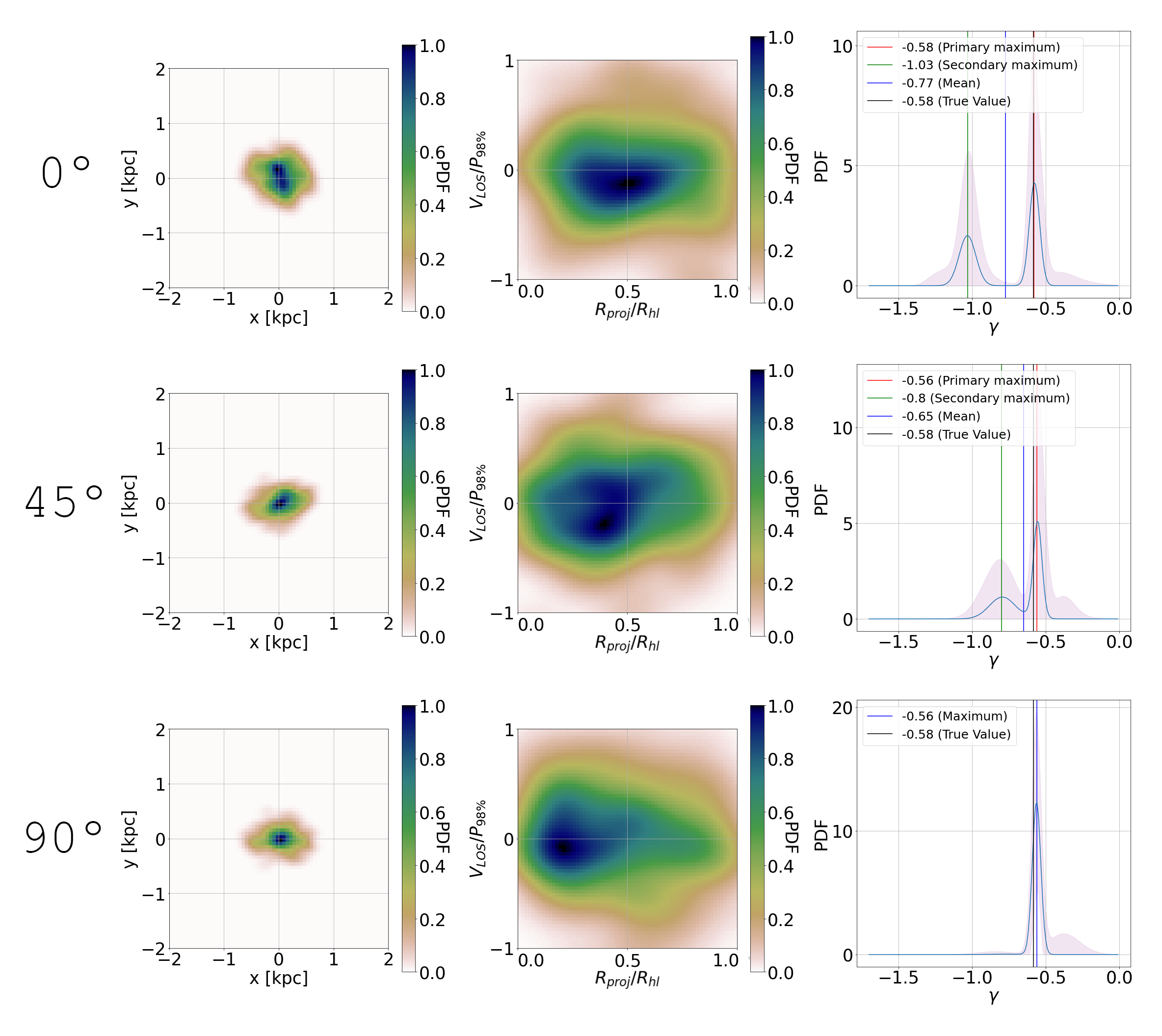}
\caption{Probability density distributions used by the neural network as  input in the case of one simulated galaxy subset seen at 0º (face-on), 45º and 90º (side-on), alongside with the Bayesian posteriors predicted by the model. Left columns: PDFs in the \{x,y\} phase space. Central columns: PDFs in the \{$\hat{R}_{\rm proj},\hat{v}_{\rm LOS}$\} phase space. Right columns: predicted Bayesian posterior in the space of inner slope of the DM profile (slope at 150 pc); shaded regions represent the standard deviation of the posterior values for the MC-Dropout inferences at each slope point. The red vertical line shows the primary maximum of the posterior distribution (the mode), the green one the secondary maximum  and the black one the true value of the inner DM slope. As a blue line, the mean between primary and secondary maximum is shown, when two peaks exist (in the bottom panel, instead, the blue line represents the unique maximum). This example shows how the appearance of double peaks in the posterior distributions is strongly related to the viewing angle.}
\label{fig:example_posteriors_2} 
\end{figure*}

Table \ref{tab:coverage} shows the percentage of the test dataset projections for which the true value of their inner slope is recovered  within different multiples of $\sigma_{\rm pos}$. If we approximate the posteriors to single Gaussians (which is a proper approximation for roughly 90\% of the projections), a well calibrated uncertainty should provide around 68\% of the outputs within a confidence level of  $1-\sigma_{\rm pos}$. Our greater percentage ($\sim86\%$) of projections within the confidence level of  $1-\sigma_{\rm pos}$ indicates that the model is over-predicting the uncertainties $\sigma_{\rm pos}$, yielding broader posteriors than it should. This can be an effect of a too high dropout rate (see \ref{sec:appendix}) during training, which has been shown to have such a outcome on the results of probabilistic neural network models \citep{ghosh22}. As it is, our model should be interpreted as conservative, since a future, better calibrated MDCNN would provide even tighter uncertainties in recovering the true inner slope of a galaxy.

\subsection{Effect of viewing angle on the inference of DM slopes}

Most of the posteriors for the different projections have an approximately normal distribution (the second Gaussian disappearing or constituting a skewness correction to the main Gaussian), but several of them have two distinct peaks. Specifically, around 30\% of the galaxies have double peaks in more than 10\% of their posteriors. 54$\%$ of these galaxies are cored  while 46\% are cuspy, indicating that the appearance of double peaks in the PDFs arises in both scenarios (here, we define as cored galaxies those with inner slope $-0.6<\gamma<0$, and cuspy any galaxy with $\gamma<-0.6$).
In Fig. \ref{fig:example_posteriors_1} and \ref{fig:example_posteriors_2} we show the PDFs and posteriors of two galaxies at different observation angles, spanning the range between a face-on and a edge-on view. 
Strikingly, these images show that the width of the PDFs as well as the appearance of double peaks are strongly related to the viewing angle of the galaxy.
This indicates that the appearance of double peaks is a consequence of the fact that some information on the underlying DM profiles is   hidden when viewing the galaxy at some particular angle, while it is released and efficiently passed to the network when looking at the galaxy  from other angles: this finding  has profound consequences for the  interpretation of `cusp-cores' in dwarfs. For example, in Fig \ref{fig:example_posteriors_2} we observe that the double peaks in the posterior distribution disappear when the galaxy is seen edge-on,  while  a face-on configuration provides a second peak that mimics the presence of a cusp. 

\begin{figure*}
\includegraphics[width=3.18in,height=2.9in]{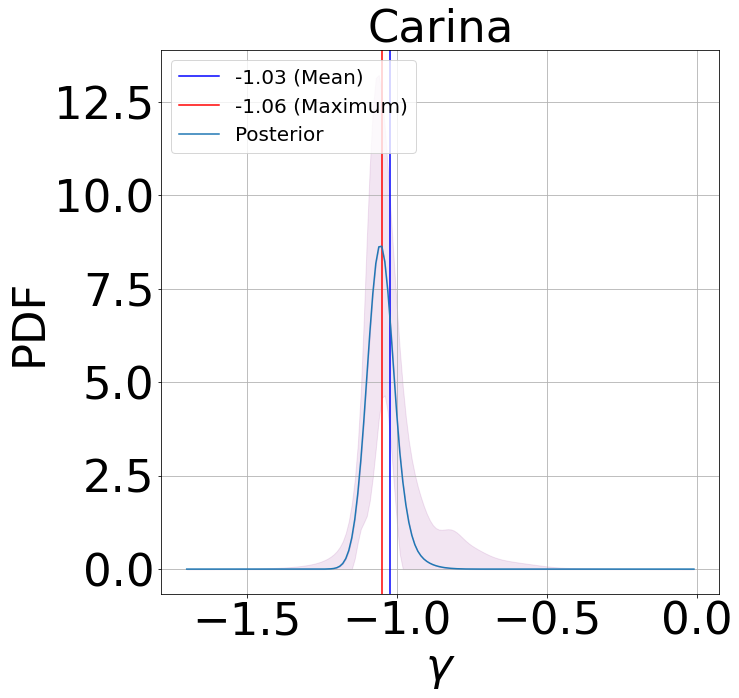}
\hspace{0.2in} \includegraphics[width=2.9in,height=2.9in]{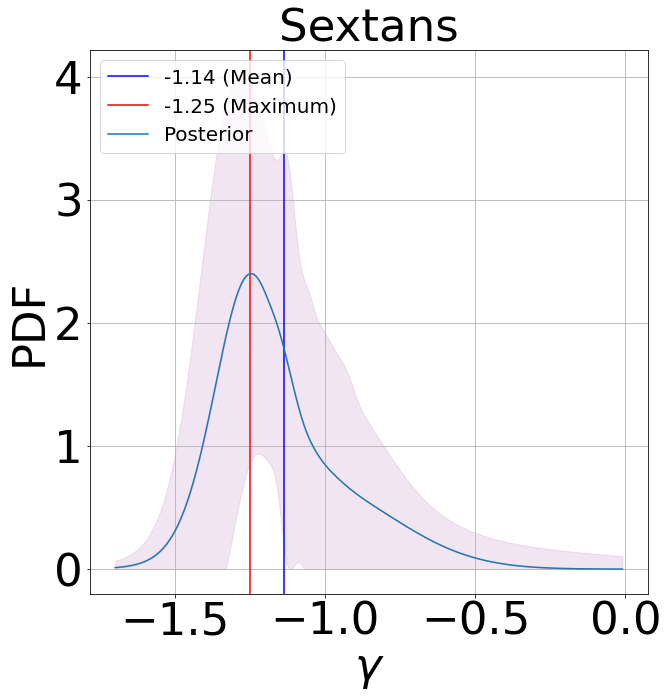}
\\
\hspace{0.1in} \includegraphics[width=3in,height=2.9in]{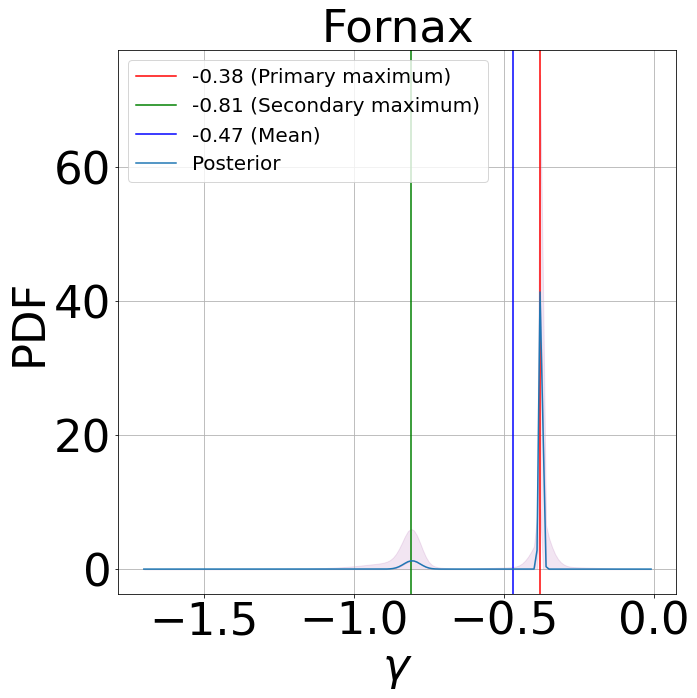}
\includegraphics[width=3.22in,height=2.9in]{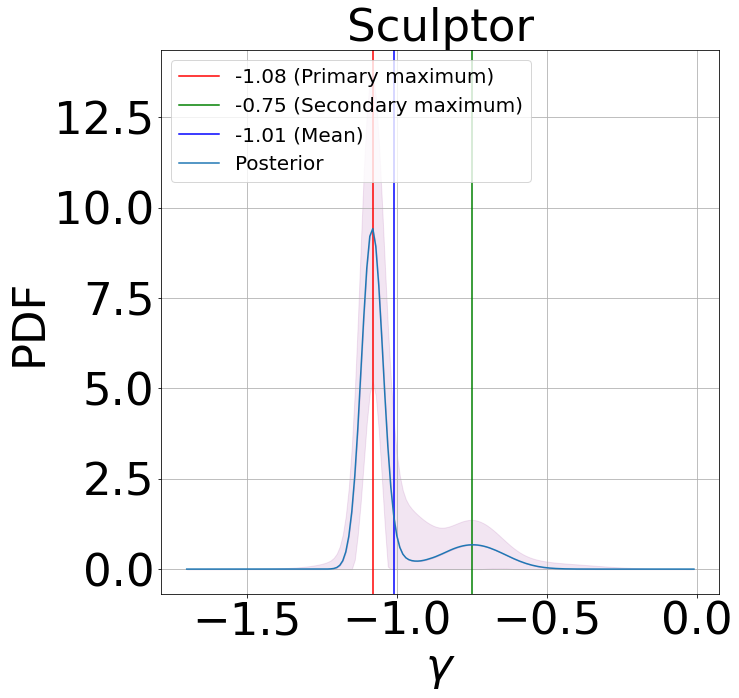}
\caption{Bayesian posterior distributions in the space of inner slope of  DM profiles (slope at 150 pc) predicted by our neural network model for the observed dSphs Carina, Sextans, Sculptor and Fornax. Shaded regions represents the standard deviation of the posterior values for the MC-Dropout inferences at each slope point. In each panel, the global maximum of the posterior distribution as well as the mean value are indicated, together with primary and secondary peaks when they exist. Fornax has the strongest signature of a central DM density core, while Carina has the strongest signature of having a NFW profile. Sextans is cuspy, though with a large uncertanty, while Sculptor is cuspy with a secondary peak indicating a mild core.}
\label{fig:obs_posteriors} 
\end{figure*}

However, this is just an example, and we have several cases of galaxies in which the double peaks appear in edge-on view and disappear in face-on, so that the appearance of these multiple peaks is not related to a specific edge-on or face-on configuration: indeed, the distribution of angles for those PDFs showing  double peaks  is uniform throughout the complete dataset. The occurrence, significance and widths of the double peaked PDFs will be explored in future works, as it goes beyond the scope of this paper.

\section{Application to observed galaxies}

We  proceed to test our model with real observed galaxies, in order to ensure the applicability of the model  and to verify that the neural network is not detecting features of simulated galaxies that do not correspond to any real physical system.
We selected  four dSphs for which detailed spectroscopic samples of stellar-kinematic data have been published. At this stage, we adopt  the catalogs by \cite{Walker09a} to directly compare our results with those obtained using the code {\sc GravSphere}, as in \cite{ReadJustin19}. The selected galaxies are
Carina, Sextans, Fornax and Sculptor, for which we further use the center position, velocity, ellipticity and half-light radius as compiled in \cite{Battaglia22}.
\\
To build our input PDFs, we considered only those stars with a 90\% or higher probability of being part of the galaxy and we took the mean value of the line-of-sight velocity for those stars with multiple measurements. In total, we considered 460 stars for Carina, 1353 for Fornax, 809 for Sculptor and 327 for Sextans, and we used their projected x-y positions and line-of-sight velocities. The x-y positions are normalized using the circularized half-light radius $R'_{\rm hlr} = R_{\rm hlr}\sqrt{1-\rm ell}$, where $R_{\rm hlr}$ and ell are the half-light radius and ellipticity from \cite{Battaglia22}.

\subsection{Deriving central DM density slopes of dSphs with CNNs}

We now infer the inner slope of the observed dwarfs. Fig. \ref{fig:obs_posteriors} shows the posterior distributions constructed by the model for each observed galaxy. Fornax presents a very narrow peak around $\gamma = -0.38$, indicating that this galaxy has a strong central DM core, while a secondary peak would give a 12\% probability that the inner slope is around $\gamma = -0.81$. This is consistent with several previous works that predict a cored profile for Fornax \cite[see][amongst others]{Goerdt06,Walker11, brook15a,Pascale18}. For the other three galaxies, a cusp is predicted with varying degrees of certainty. The model has a clear peak around $\gamma = -1.06$ for Carina, which roughly corresponds to the slope of a NFW profile at 150 pc. 

\begin{table}
\label{tab:masses}
\begin{center}
\begin{tabular}{lcc}
\hline
\hline
 & $\gamma_{GS}$ & $\gamma_{NN}$\\
\hline
Carina   & $-1.23^{+0.39}_{-0.35}$ & $-1.06^{+0.05}_{-0.04}$   \\                   
Sextans  & $-0.95^{+0.25}_{-0.25}$ & $-1.25^{+0.25}_{-0.09}$ \\ 
Fornax   & $-0.30^{+0.21}_{-0.28}$ & $-0.38^{+0.01}_{-0.02}$   \\                   
Sculptor  & $-0.83^{+0.30}_{-0.25}$ & $-1.08^{+0.08}_{-0.04}$ \\ 
\hline
\hline
\end{tabular} \\
\caption{Inner slope of the DM profile (at 150 pc) for Carina, Sextans, Sculptor and Fornax galaxies predicted by {\sc GravSphere} ($\gamma_{GS}$) and our neural network ($\gamma_{NN}$), with their 68 per cent confidence intervals (for the neural network posterior, taking the primary maximum as reference).The agreement between the two methods is encouraging.}
\label{tab:comparison}
\end{center}
\end{table}

Sextans presents a relatively large uncertainty in the inner slope value, as depicted by the quite broad PDFs, with a broad peak around $\gamma = -1.25$ and a strong right wing that does not fall below 10$\%$ of the peak value  until it reaches $\gamma = -0.68$. 
Finally,  Sculptor peaks at $\gamma = -1.08$, but it has a wide secondary peak, predicting a 18\% probability of having a mild core with  $\gamma = -0.75$. A small core was derived for Sculptor by using kinematical data and a mass-dependent profile fit in \citet{brook15a}, in agreement with the \citet{Walker11} and \citet{Agnello12} methods that, employing multiple stellar populations within a galaxy, also  predicted a core in such dwarf (see also \citealt{Zhu16,Breddels13a,Hayashi20}). Other  studies, however, surprisingly predict a cusp for Sculptor after all \citep{richardson14}, highlighting the importance of deriving the DM density of this dSphs with several different methods.
Our derived posterior distributions offer great versatility in interpreting the results, allowing for a more complex analysis compared to models that only allow for uncertainty ranges around the inferred value.

We compare the results of our model with the inner slopes inferred for these same galaxies at 150 pc using  {\sc GravSphere}, a non-parametric spherical Jeans analysis code, which make use of photometric and kinematic data from the galaxies \citep{ReadJustin19}. The inferred values, along with their 68 per cent confidence intervals (in our case, taking the primary maximum as reference), are listed in Table \ref{tab:comparison}. The derived values are consistent between the two models,  within their respective uncertainty ranges, indicating that our neural network model is making predictions similar to those obtained by Jeans analysis. Furthermore, the accuracy of our neural network is greater, with errors roughly an order of magnitude smaller than those of {\sc GravSphere}: this preliminary finding will be expanded and explored in more detail in  future work.

\begin{figure}
\includegraphics[width=\columnwidth]{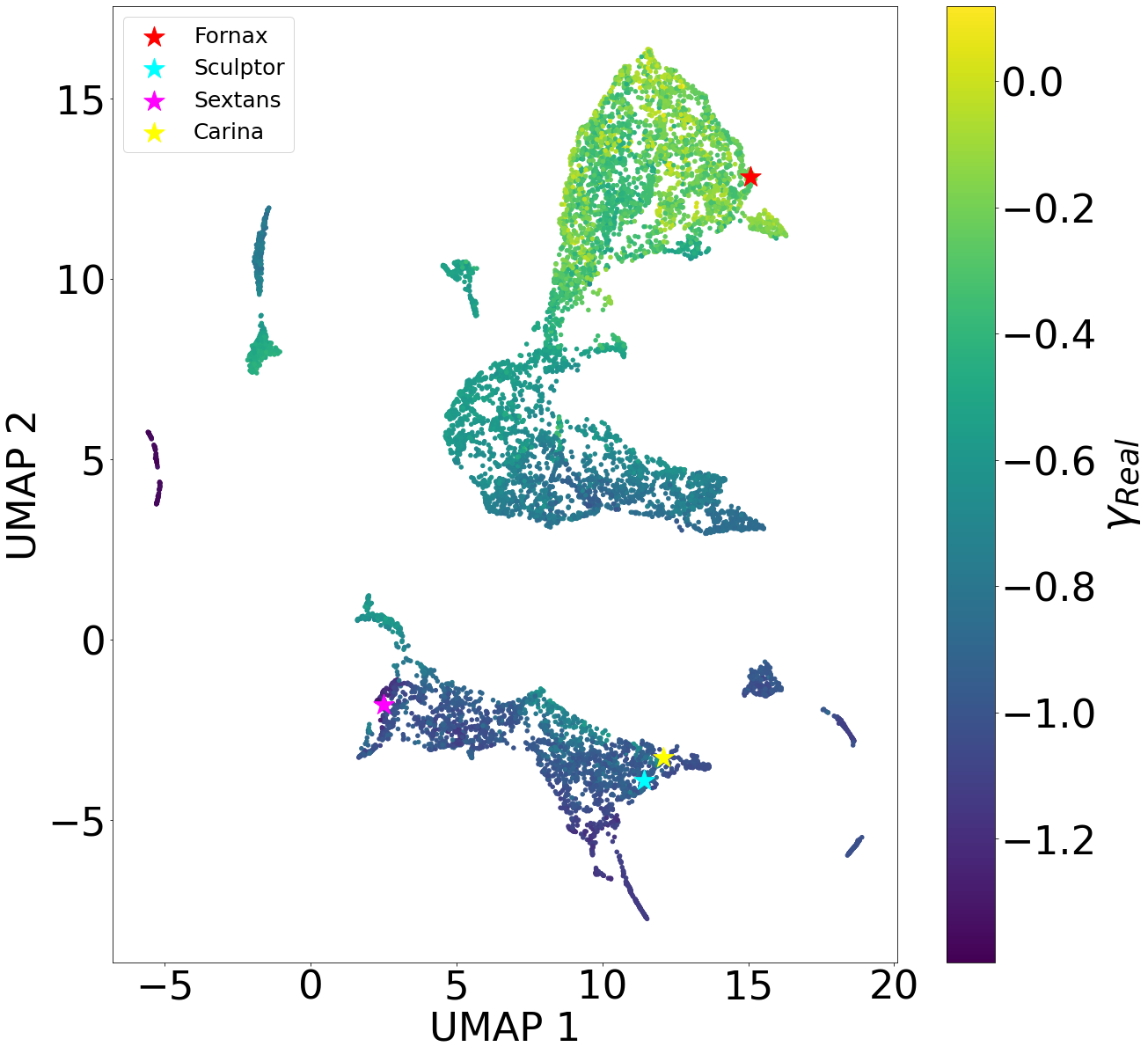}
\caption{Representation of the parameter space for the test data from simulated galaxies reduced to 2-dimension with a  Uniform Manifold Approximation and Projection (UMAP), color coded by the real expected inner DM slope. Plotted as colored stars are the locations in the reduced parameter space of Carina, Sextans, Sculptor and Fornax galaxies. Note that different inner slopes occupy different areas of the plot and, importantly,  observed dwarf galaxies fall well within   the simulation region,   indicating that the neural network model is not seeing relevant differences between the simulated data with which we have fed it and the observational data.}
\label{fig:param_space} 
\end{figure}

Compared to {\sc GravSphere} and similar codes, the neural network approach is significantly faster. In a modern laptop, {\sc GravSphere} will need about half a day to run an analysis of one of these galaxies, whereas the neural network can be trained with the amount of data used in this work in less than half an hour on a standard GPU. 
Furthermore, the training and the evaluation are independent calculation in a neural network model, which means that, once the model has been trained, its application to any input data to construct the posterior distribution is nearly instantaneous. This feature will not change no matter how much the model is expanded and complexified to perform more complete analyses of the galaxy of interest.

\subsection{Testing the similarity of training vs observational data}

When training a neural network with simulations to then perform inference on real data, there is always the risk that the  network will detect and learn from specific features of the simulation code that do not correspond to reality, and  this would cause issues when interpreting observational data, as they have different qualities than those used in the training set.
We can test the degree to which our network sees observational data as equivalent to the data it has been trained on by observing the parameter space  of the test dataset, defined  as the set of all combinations of the six parameters corresponding to each element of such dataset.
Namely, our outputs are defined by the mean, standard deviation and weight of two Gaussians: this 6D parameter space will have regions populated with points and regions completely empty,  corresponding to the combinations of parameters that do not parameterise the characteristics of any physical system found in the dataset.
If the neural network does not see differences in the input with respect to the data it has been trained on, the resulting parameters, coming from the evaluation of observational data with our model, will fall within the populated regions of the parameter space of the simulation dataset.

To be able to visualize the 6D parameter space and test if this is the case, we use the Uniform Manifold Approximation and Projection for Dimension Reduction (UMAP) technique from \cite{McInnes18} to perform a dimension reduction from 6D to 2D, thus mapping each combination of means, standard deviations and weights to only two adimensional parameters representing such `contraction', preserving the global structure of the original parameter space. This allows to visualize the parameter space in 2D. The result of the dimension reduction process from the complete test sample can be seen in Fig. \ref{fig:param_space}, alongside with the position of the four observed dwarf galaxies shown in the same parameter space, each indicated as colored star. 
As expected, the spatial location of the points in the parameter space is strongly linked to the value of their inner slope: points with a similar inner slope cluster together, showing that the network is properly parameterizing the inner slope of  galaxies during training. 
Interestingly, the four observed galaxies fall into the regions occupied by the simulated ones, which indicates that the model is considering them as data of equivalent nature as the test data. However, the fact that all four are close to the edges of the simulation input parameters could indicate the presence of some features that the model has not found in the simulations. 
The possible causes of this will be explored in future work employing a larger  observational sample.

\section{Conclusions} \label{sec:conclusions}

We present a novel model for determining the slope of the inner density profile of dark matter (DM) halos with robust uncertainty quantification using machine learning techniques. The goal of this work is to be able to infer such density slopes ($\gamma$)  by simply using positions and velocities of  stars within  galaxies.
Our method uses mixture density convolutional neural networks with a Gaussian density layer backend to model complex galaxy substructure. We use  line-of-sight velocities and positions of stars projected on the sky within simulated dwarf galaxies, employing Kernel Density Estimations (KDEs) to construct continuous 2D probability density functions (PDFs) of the distribution of such stars in \{$\hat{R}_{\rm proj},\hat{v}_{\rm LOS}$\}  and  \{x,y\} phase space, which serve as input to our neural network using a double channel architecture (Fig.\ref{fig:example_inputs} and \ref{fig:architecture}).

We train and evaluate our model using a large set of fully cosmological simulations of dwarf galaxies with halo masses of $10^9$ to $10^{11.5}$ \msun, and stellar masses of $10^5$  to $10^{9.5}$ \msun, from the NIHAO and AURIGA projects \citep{wang15,dutton20,Grand17}. The use of different physical models employed in these simulations allows us to have a range of  density profiles at each particular galaxy mass, including both cores and cusps (Fig.\ref{fig:gamma-M}). All simulated galaxies and their subsets of stars are
randomly projected in several sky planes, to simulate several viewing angles.

The loss function to minimize during the training is the negative logarithmic likelihood of the training sample, defined as a multivariate Gaussian probability distribution, which is the output of our Gaussian density layer backend. This allows a flexible probabilistic representation of the results, which yields accurate and statistically consistent uncertainties. 
For each galaxy the neural network outputs a PDF which gives the posterior probability
of a certain slope to be the inner slope of the galaxy.

The main results of this work are listed here:

\begin{itemize}
    \item the inner slope of simulated galaxies is predicted with a mean absolute deviation of  $\mu_{\epsilon} =0.056$ (where the deviation is defined as $\epsilon$=$\gamma_{\rm Real}-\gamma_{\rm Pred}$, and the predicted inner slope,  $\gamma_{\rm Pred}$, is obtained from the mode of the PDFs) and a standard deviation of $\sigma_\epsilon = 0.075$ for the whole sample (Fig. \ref{fig:acc_error} and \ref{fig:hist_error});
    \item  82$\%$ (98$\%$) of the  galaxies have their inner slope correctly determined within $\pm$ 0.1 (0.3) of their true value (Table \ref{tab:ac_error});
    \item  the posteriors PDFs have a mean standard deviation of $\sigma_{\rm pos} = 0.108$,  showing no bias towards more accuracy for cuspy or cored galaxies (Fig.\ref{fig:sigmas});
    \item while in most cases the output of the model is a single peaked PDF, in $\sim 30\%$ of the galaxies some of their projections show a double peak: we demonstrated that this is related to some  viewing angles, indicating the importance of properly determining the inclination of galaxies (Fig. \ref{fig:example_posteriors_1} and \ref{fig:example_posteriors_2});
    \item when applied to a set of four observed dSphs, our model recovers their inner slopes  
    yielding values consistent with those obtained with  the Jeans modelling based code  {\sc GravSphere}  as in  \citet{ReadJustin19} (Table \ref{tab:comparison});
    \item we found that the Fornax
dSph has a strong indication of having a central DM core, Carina and Sextans have cusps
(although the latter with a large uncertainty), while Sculptor shows a double peaked PDF
indicating that a cusp is preferred, but a core can not be ruled out (Fig.\ref{fig:obs_posteriors}). These results are in agreement with several previously derived inner slopes for these galaxies.
\end{itemize}

The current architecture could be used as a basis for building models that provide a more complete output, such as a  prediction of the full density profile of galaxies. The nature of the neural network allows it to be constantly extended and improved. While we have implemented a network of relatively low complexity, there are a series of interesting possibilities with a further level of sophistication that are worth exploring. For example, the use of normalizing flows may yield to more robust results \citep{ramanah20b} while the use of a 3D convolutional network applied to PDFs defined in the \{x,y,$\hat{v}_{\rm LOS}$\} phase space has given good results in  galaxy cluster masses inference \citep{ramanah20a}. 

In the future, the architecture of this model could  be expanded by including more input data, such as  surface brightness profiles or proper motion of  stars from missions like \textit{GAIA} \citep{Gaia22}. Furthermore, the inclusion of other spectroscopic samples present in the literature, as well as of those soon to be acquired with upcoming facilities, will  certainly be beneficial for this analysis.
Adapting the architecture and  introducing more information may enable the network to improve accuracy and reduce the range of variability of the results with respect to the angle of observation, an avenue that will be explored in future works. 

We  have shown that deep  learning techniques provide a innovative  method for the determination of the inner DM profiles in dwarf galaxies, complementary to the use of Jeans and Schwarzschild modelling, achieving great accuracy and offering a  complex representation of uncertainties.

Our newly developed neural network method is a promising tool for the study of the mass distribution within dwarf galaxies, which in turn can help  discriminate between different  models and, in such, constraining the properties of the elusive DM.

\section*{Acknowledgements}
Author contribution:
C.B.B. led the project.
J.E.M. performed the analysis. 
J.E.M. and A.D.C. wrote the manuscript.
J.E.M. and M.H.C. developed the machine learning architecture.
C.B.B., M.H.C. and A.D.C. supervised the student during the project. A.V.M. and R.J.J.G. provided the simulation data.
G.B. helped with the selection of the observational dataset. All authors provided feedback on the draft.

CB is supported by the Spanish Ministry of Science and Innovation (MICIU/FEDER)
through research grant PID2021-122603NB-C22. ADC is supported by a \textit{Junior Leader fellowship} from `La Caixa' Foundation (ID 100010434), code  LCF/BQ/PR20/11770010. She further acknowledges Macquarie University for the hospitality during the preparation of this work as a  Honorary Visiting Fellow. G.B. acknowledges support from the Agencia Estatal de Investigación del Ministerio de Ciencia en Innovación (AEI-MICIN) under grant references PID2020-118778GB-I00/10.13039/501100011033 and  grant number CEX2019-000920-S. Fellow R.G. acknowledges financial support from the Spanish Ministry of Science and Innovation (MICINN) through the Spanish State Research Agency, under the Severo Ochoa Program 2020-2023 (CEX2019-000920-S). Part of this research was carried out on the High Performance Computing resources at New York University Abu Dhabi (UAE).

\section*{Data Availability}
The data used in this work are available upon reasonable request to the
corresponding author and to the PIs of the NIHAO and AURIGA projects.
\vspace{.0cm}\bibliographystyle{mn2e}
\bibliography{archive}

\begin{thebibliography}{}
\makeatletter
\relax
\def\mn@urlcharsother{\let\do\@makeother \do\$\do\&\do\#\do\^\do\_\do\%\do\~}
\def\mn@doi{\begingroup\mn@urlcharsother \@ifnextchar [ {\mn@doi@}
  {\mn@doi@[]}}
\def\mn@doi@[#1]#2{\def\@tempa{#1}\ifx\@tempa\@empty \href
  {http://dx.doi.org/#2} {doi:#2}\else \href {http://dx.doi.org/#2} {#1}\fi
  \endgroup}
\def\mn@eprint#1#2{\mn@eprint@#1:#2::\@nil}
\def\mn@eprint@arXiv#1{\href {http://arxiv.org/abs/#1} {{\tt arXiv:#1}}}
\def\mn@eprint@dblp#1{\href {http://dblp.uni-trier.de/rec/bibtex/#1.xml}
  {dblp:#1}}
\def\mn@eprint@#1:#2:#3:#4\@nil{\def\@tempa {#1}\def\@tempb {#2}\def\@tempc
  {#3}\ifx \@tempc \@empty \let \@tempc \@tempb \let \@tempb \@tempa \fi \ifx
  \@tempb \@empty \def\@tempb {arXiv}\fi \@ifundefined
  {mn@eprint@\@tempb}{\@tempb:\@tempc}{\expandafter \expandafter \csname
  mn@eprint@\@tempb\endcsname \expandafter{\@tempc}}}

\bibitem[\protect\citeauthoryear{{Agnello} \& {Evans}}{{Agnello} \&
  {Evans}}{2012}]{Agnello12}
{Agnello} A.,  {Evans} N.~W.,  2012, \mn@doi [\apjl]
  {10.1088/2041-8205/754/2/L39}, \href
  {https://ui.adsabs.harvard.edu/abs/2012ApJ...754L..39A} {754, L39}

\bibitem[\protect\citeauthoryear{{Battaglia}, {Helmi}, {Tolstoy}, {Irwin},
  {Hill}  \& {Jablonka}}{{Battaglia} et~al.}{2008}]{battaglia08}
{Battaglia} G.,  {Helmi} A.,  {Tolstoy} E.,  {Irwin} M.,  {Hill} V.,
  {Jablonka} P.,  2008, \mn@doi [\apjl] {10.1086/590179}, \href
  {http://adsabs.harvard.edu/abs/2008ApJ...681L..13B} {681, L13}

\bibitem[\protect\citeauthoryear{{Battaglia}, {Taibi}, {Thomas}  \&
  {Fritz}}{{Battaglia} et~al.}{2022}]{Battaglia22}
{Battaglia} G.,  {Taibi} S.,  {Thomas} G.~F.,   {Fritz} T.~K.,  2022, \mn@doi
  [\aap] {10.1051/0004-6361/202141528}, \href
  {https://ui.adsabs.harvard.edu/abs/2022A&A...657A..54B} {657, A54}

\bibitem[\protect\citeauthoryear{{Ben{\'\i}tez-Llambay}, {Frenk}, {Ludlow}  \&
  {Navarro}}{{Ben{\'\i}tez-Llambay} et~al.}{2019}]{Benitez19}
{Ben{\'\i}tez-Llambay} A.,  {Frenk} C.~S.,  {Ludlow} A.~D.,   {Navarro} J.~F.,
  2019, \mn@doi [\mnras] {10.1093/mnras/stz1890}, \href
  {https://ui.adsabs.harvard.edu/abs/2019MNRAS.488.2387B} {488, 2387}

\bibitem[\protect\citeauthoryear{{Binney} \& {Mamon}}{{Binney} \&
  {Mamon}}{1982}]{Binney82}
{Binney} J.,  {Mamon} G.~A.,  1982, \mn@doi [\mnras] {10.1093/mnras/200.2.361},
  \href {https://ui.adsabs.harvard.edu/abs/1982MNRAS.200..361B} {200, 361}

\bibitem[\protect\citeauthoryear{{Breddels, Maarten A.} \& {Helmi,
  Amina}}{{Breddels, Maarten A.} \& {Helmi, Amina}}{2013}]{Breddels13b}
{Breddels, Maarten A.} {Helmi, Amina} 2013, \mn@doi [A\&A]
  {10.1051/0004-6361/201321606}, 558, A35

\bibitem[\protect\citeauthoryear{Breddels, Helmi, van~den Bosch, van~de Ven  \&
  Battaglia}{Breddels et~al.}{2013}]{Breddels13a}
Breddels M.~A.,  Helmi A.,  van~den Bosch R. C.~E.,  van~de Ven G.,   Battaglia
  G.,  2013, \mn@doi [Monthly Notices of the Royal Astronomical Society]
  {10.1093/mnras/stt956}, 433, 3173

\bibitem[\protect\citeauthoryear{{Brook} \& {Di Cintio}}{{Brook} \& {Di
  Cintio}}{2015}]{brook15a}
{Brook} C.~B.,  {Di Cintio} A.,  2015, \mn@doi [\mnras] {10.1093/mnras/stv864},
  \href {http://adsabs.harvard.edu/abs/2015MNRAS.450.3920B} {450, 3920}

\bibitem[\protect\citeauthoryear{{Bullock} \& {Boylan-Kolchin}}{{Bullock} \&
  {Boylan-Kolchin}}{2017}]{bullock17}
{Bullock} J.~S.,  {Boylan-Kolchin} M.,  2017, \mn@doi [\araa]
  {10.1146/annurev-astro-091916-055313}, \href
  {https://ui.adsabs.harvard.edu/abs/2017ARA&A..55..343B} {55, 343}

\bibitem[\protect\citeauthoryear{{Cappellari} et~al.,}{{Cappellari}
  et~al.}{2006}]{cappellari06}
{Cappellari} M.,  et~al., 2006, \mn@doi [\mnras]
  {10.1111/j.1365-2966.2005.09981.x}, \href
  {https://ui.adsabs.harvard.edu/abs/2006MNRAS.366.1126C} {366, 1126}

\bibitem[\protect\citeauthoryear{{Chan}, {Kere{\v s}}, {O{\~n}orbe}, {Hopkins},
  {Muratov}, {Faucher-Gigu{\`e}re}  \& {Quataert}}{{Chan}
  et~al.}{2015}]{chan15}
{Chan} T.~K.,  {Kere{\v s}} D.,  {O{\~n}orbe} J.,  {Hopkins} P.~F.,  {Muratov}
  A.~L.,  {Faucher-Gigu{\`e}re} C.-A.,   {Quataert} E.,  2015, \mn@doi [\mnras]
  {10.1093/mnras/stv2165}, \href
  {http://adsabs.harvard.edu/abs/2015MNRAS.454.2981C} {454, 2981}

\bibitem[\protect\citeauthoryear{{Collins} et~al.,}{{Collins}
  et~al.}{2021}]{Collins21}
{Collins} M. L.~M.,  et~al., 2021, \mn@doi [\mnras] {10.1093/mnras/stab1624},
  \href {https://ui.adsabs.harvard.edu/abs/2021MNRAS.505.5686C} {505, 5686}

\bibitem[\protect\citeauthoryear{{Di Cintio}, {Brook}, {Macci{\`o}}, {Stinson},
  {Knebe}, {Dutton}  \& {Wadsley}}{{Di Cintio} et~al.}{2014a}]{DiCintio2014a}
{Di Cintio} A.,  {Brook} C.~B.,  {Macci{\`o}} A.~V.,  {Stinson} G.~S.,  {Knebe}
  A.,  {Dutton} A.~A.,   {Wadsley} J.,  2014a, \mn@doi [\mnras]
  {10.1093/mnras/stt1891}, \href
  {http://adsabs.harvard.edu/abs/2014MNRAS.437..415D} {437, 415}

\bibitem[\protect\citeauthoryear{{Di Cintio}, {Brook}, {Dutton}, {Macci{\`o}},
  {Stinson}  \& {Knebe}}{{Di Cintio} et~al.}{2014b}]{DiCintio2014b}
{Di Cintio} A.,  {Brook} C.~B.,  {Dutton} A.~A.,  {Macci{\`o}} A.~V.,
  {Stinson} G.~S.,   {Knebe} A.,  2014b, \mn@doi [\mnras]
  {10.1093/mnras/stu729}, \href
  {http://adsabs.harvard.edu/abs/2014MNRAS.441.2986D} {441, 2986}

\bibitem[\protect\citeauthoryear{{Dutton}, {Buck}, {Macci{\`o}}, {Dixon},
  {Blank}  \& {Obreja}}{{Dutton} et~al.}{2020}]{dutton20}
{Dutton} A.~A.,  {Buck} T.,  {Macci{\`o}} A.~V.,  {Dixon} K.~L.,  {Blank} M.,
  {Obreja} A.,  2020, \mn@doi [\mnras] {10.1093/mnras/staa3028}, \href
  {https://ui.adsabs.harvard.edu/abs/2020MNRAS.499.2648D} {499, 2648}

\bibitem[\protect\citeauthoryear{{Gaia Collaboration} et~al.,}{{Gaia
  Collaboration} et~al.}{2021}]{Gaia22}
{Gaia Collaboration} et~al., 2021, \mn@doi [\aap]
  {10.1051/0004-6361/202039657}, \href
  {https://ui.adsabs.harvard.edu/abs/2021A&A...649A...1G} {649, A1}

\bibitem[\protect\citeauthoryear{Gal \& Ghahramani}{Gal \&
  Ghahramani}{2015}]{Gal15}
Gal Y.,  Ghahramani Z.,  2015, Proceedings of The 33rd International Conference
  on Machine Learning

\bibitem[\protect\citeauthoryear{Geha, Guhathakurta, Rich  \& Cooper}{Geha
  et~al.}{2006}]{Geha06}
Geha M.~C.,  Guhathakurta P.,  Rich R.~M.,   Cooper M.~C.,  2006, The
  Astronomical Journal, 131, 332

\bibitem[\protect\citeauthoryear{{Gentile}, {Salucci}, {Klein}, {Vergani}  \&
  {Kalberla}}{{Gentile} et~al.}{2004}]{gentile04}
{Gentile} G.,  {Salucci} P.,  {Klein} U.,  {Vergani} D.,   {Kalberla} P.,
  2004, \mn@doi [\mnras] {10.1111/j.1365-2966.2004.07836.x}, \href
  {https://ui.adsabs.harvard.edu/abs/2004MNRAS.351..903G} {351, 903}

\bibitem[\protect\citeauthoryear{Ghosh et~al.,}{Ghosh et~al.}{2022}]{ghosh22}
Ghosh A.,  et~al., 2022, GaMPEN: A Machine Learning Framework for Estimating
  Bayesian Posteriors of Galaxy Morphological Parameters,
  \mn@doi{10.48550/ARXIV.2207.05107}, \url {https://arxiv.org/abs/2207.05107}

\bibitem[\protect\citeauthoryear{Gnedin \& Zhao}{Gnedin \&
  Zhao}{2002}]{gnedin02}
Gnedin O.~Y.,  Zhao H.,  2002, \mn@doi [Monthly Notices of the Royal
  Astronomical Society] {10.1046/j.1365-8711.2002.05361.x}, 333, 299

\bibitem[\protect\citeauthoryear{Goerdt, Moore, Read, Stadel  \& Zemp}{Goerdt
  et~al.}{2006}]{Goerdt06}
Goerdt T.,  Moore B.,  Read J.~I.,  Stadel J.,   Zemp M.,  2006, \mn@doi
  [Monthly Notices of the Royal Astronomical Society]
  {10.1111/j.1365-2966.2006.10182.x}, 368, 1073

\bibitem[\protect\citeauthoryear{{Governato} et~al.,}{{Governato}
  et~al.}{2010}]{governato10}
{Governato} F.,  et~al., 2010, \mn@doi [Nature] {10.1038/nature08640}, \href
  {http://adsabs.harvard.edu/abs/2010Natur.463..203G} {463, 203}

\bibitem[\protect\citeauthoryear{Grand et~al.,}{Grand et~al.}{2017}]{Grand17}
Grand R. J.~J.,  et~al., 2017, \mn@doi [Monthly Notices of the Royal
  Astronomical Society] {10.1093/mnras/stx071}, 467, 179

\bibitem[\protect\citeauthoryear{Hayashi, Chiba  \& Ishiyama}{Hayashi
  et~al.}{2020}]{Hayashi20}
Hayashi K.,  Chiba M.,   Ishiyama T.,  2020, \mn@doi [The Astrophysical
  Journal] {10.3847/1538-4357/abbe0a}, 904, 45

\bibitem[\protect\citeauthoryear{Ho, Rau, Ntampaka, Farahi, Trac  \&
  P{\'{o}}czos}{Ho et~al.}{2019}]{ho19}
Ho M.,  Rau M.~M.,  Ntampaka M.,  Farahi A.,  Trac H.,   P{\'{o}}czos B.,
  2019, \mn@doi [The Astrophysical Journal] {10.3847/1538-4357/ab4f82}, 887, 25

\bibitem[\protect\citeauthoryear{{Jaffe}}{{Jaffe}}{1983}]{Jaffe83}
{Jaffe} W.,  1983, \mn@doi [\mnras] {10.1093/mnras/202.4.995}, \href
  {https://ui.adsabs.harvard.edu/abs/1983MNRAS.202..995J} {202, 995}

\bibitem[\protect\citeauthoryear{Kaplinghat, Tulin  \& Yu}{Kaplinghat
  et~al.}{2016}]{kaplinghat16}
Kaplinghat M.,  Tulin S.,   Yu H.-B.,  2016, \mn@doi [Phys. Rev. Lett.]
  {10.1103/PhysRevLett.116.041302}, 116, 041302

\bibitem[\protect\citeauthoryear{Kingma \& Ba}{Kingma \& Ba}{2014}]{Diederik14}
Kingma D.~P.,  Ba J.,  2014, Adam: A Method for Stochastic Optimization,
  \mn@doi{10.48550/ARXIV.1412.6980}, \url {https://arxiv.org/abs/1412.6980}

\bibitem[\protect\citeauthoryear{{Kleyna}, {Wilkinson}, {Evans}  \&
  {Gilmore}}{{Kleyna} et~al.}{2001}]{kleyna01}
{Kleyna} J.~T.,  {Wilkinson} M.~I.,  {Evans} N.~W.,   {Gilmore} G.,  2001,
  \mn@doi [\apjl] {10.1086/338603}, \href
  {https://ui.adsabs.harvard.edu/abs/2001ApJ...563L.115K} {563, L115}

\bibitem[\protect\citeauthoryear{{Kodi Ramanah}, {Wojtak}, {Ansari}, {Gall}  \&
  {Hjorth}}{{Kodi Ramanah} et~al.}{2020}]{ramanah20b}
{Kodi Ramanah} D.,  {Wojtak} R.,  {Ansari} Z.,  {Gall} C.,   {Hjorth} J.,
  2020, \mn@doi [\mnras] {10.1093/mnras/staa2886}, \href
  {https://ui.adsabs.harvard.edu/abs/2020MNRAS.499.1985K} {499, 1985}

\bibitem[\protect\citeauthoryear{{Kodi Ramanah}, {Wojtak}  \& {Arendse}}{{Kodi
  Ramanah} et~al.}{2021}]{ramanah20a}
{Kodi Ramanah} D.,  {Wojtak} R.,   {Arendse} N.,  2021, \mn@doi [\mnras]
  {10.1093/mnras/staa3922}, \href
  {https://ui.adsabs.harvard.edu/abs/2021MNRAS.501.4080K} {501, 4080}

\bibitem[\protect\citeauthoryear{{Kowalczyk}, {{\L}okas}  \&
  {Valluri}}{{Kowalczyk} et~al.}{2017}]{Kowalczyk17}
{Kowalczyk} K.,  {{\L}okas} E.~L.,   {Valluri} M.,  2017, \mn@doi [\mnras]
  {10.1093/mnras/stx1520}, \href
  {https://ui.adsabs.harvard.edu/abs/2017MNRAS.470.3959K} {470, 3959}

\bibitem[\protect\citeauthoryear{LeCun, Bengio  \& Hinton}{LeCun
  et~al.}{2015}]{lecun15}
LeCun Y.,  Bengio Y.,   Hinton G.,  2015, nature, 521, 436

\bibitem[\protect\citeauthoryear{{Lelli}, {McGaugh}  \& {Schombert}}{{Lelli}
  et~al.}{2016}]{lelli16}
{Lelli} F.,  {McGaugh} S.~S.,   {Schombert} J.~M.,  2016, \mn@doi [\aj]
  {10.3847/0004-6256/152/6/157}, \href
  {https://ui.adsabs.harvard.edu/abs/2016AJ....152..157L} {152, 157}

\bibitem[\protect\citeauthoryear{{Macci{\`o}}, {Crespi}, {Blank}  \&
  {Kang}}{{Macci{\`o}} et~al.}{2020}]{maccio20}
{Macci{\`o}} A.~V.,  {Crespi} S.,  {Blank} M.,   {Kang} X.,  2020, \mn@doi
  [\mnras] {10.1093/mnrasl/slaa058}, \href
  {https://ui.adsabs.harvard.edu/abs/2020MNRAS.495L..46M} {495, L46}

\bibitem[\protect\citeauthoryear{McInnes, Healy  \& Melville}{McInnes
  et~al.}{2018}]{McInnes18}
McInnes L.,  Healy J.,   Melville J.,  2018, UMAP: Uniform Manifold
  Approximation and Projection for Dimension Reduction,
  \mn@doi{10.48550/ARXIV.1802.03426}, \url {https://arxiv.org/abs/1802.03426}

\bibitem[\protect\citeauthoryear{{Merritt}, {Graham}, {Moore}, {Diemand}  \&
  {Terzi{\'c}}}{{Merritt} et~al.}{2006}]{Merritt06}
{Merritt} D.,  {Graham} A.~W.,  {Moore} B.,  {Diemand} J.,   {Terzi{\'c}} B.,
  2006, \mn@doi [\aj] {10.1086/508988}, \href
  {http://adsabs.harvard.edu/abs/2006AJ....132.2685M} {132, 2685}

\bibitem[\protect\citeauthoryear{{Moore}}{{Moore}}{1994}]{moore94}
{Moore} B.,  1994, \mn@doi [\nat] {10.1038/370629a0}, \href
  {http://adsabs.harvard.edu/abs/1994Natur.370..629M} {370, 629}

\bibitem[\protect\citeauthoryear{{Navarro}, {Eke}  \& {Frenk}}{{Navarro}
  et~al.}{1996a}]{navarro96b}
{Navarro} J.~F.,  {Eke} V.~R.,   {Frenk} C.~S.,  1996a, \mnras, \href
  {http://adsabs.harvard.edu/abs/1996MNRAS.283L..72N} {283, L72}

\bibitem[\protect\citeauthoryear{{Navarro}, {Frenk}  \& {White}}{{Navarro}
  et~al.}{1996b}]{navarro96}
{Navarro} J.~F.,  {Frenk} C.~S.,   {White} S.~D.~M.,  1996b, ApJ, \href
  {http://adsabs.harvard.edu/cgi-bin/nph-bib_query?bibcode=1996ApJ...462..563N&db_key=AST}
  {462, 563}

\bibitem[\protect\citeauthoryear{{Pascale}, {Posti}, {Nipoti}  \&
  {Binney}}{{Pascale} et~al.}{2018}]{Pascale18}
{Pascale} R.,  {Posti} L.,  {Nipoti} C.,   {Binney} J.,  2018, \mn@doi [\mnras]
  {10.1093/mnras/sty1860}, \href
  {https://ui.adsabs.harvard.edu/abs/2018MNRAS.480..927P} {480, 927}

\bibitem[\protect\citeauthoryear{{Pontzen} \& {Governato}}{{Pontzen} \&
  {Governato}}{2012}]{pontzen12}
{Pontzen} A.,  {Governato} F.,  2012, \mn@doi [\mnras]
  {10.1111/j.1365-2966.2012.20571.x}, \href
  {http://adsabs.harvard.edu/abs/2012MNRAS.421.3464P} {421, 3464}

\bibitem[\protect\citeauthoryear{{Read}, {Wilkinson}, {Evans}, {Gilmore}  \&
  {Kleyna}}{{Read} et~al.}{2006}]{read06}
{Read} J.~I.,  {Wilkinson} M.~I.,  {Evans} N.~W.,  {Gilmore} G.,   {Kleyna}
  J.~T.,  2006, \mn@doi [\mnras] {10.1111/j.1365-2966.2005.09959.x}, \href
  {http://adsabs.harvard.edu/abs/2006MNRAS.367..387R} {367, 387}

\bibitem[\protect\citeauthoryear{{Read}, {Walker}  \& {Steger}}{{Read}
  et~al.}{2019}]{ReadJustin19}
{Read} J.~I.,  {Walker} M.~G.,   {Steger} P.,  2019, \mn@doi [\mnras]
  {10.1093/mnras/sty3404}, \href
  {https://ui.adsabs.harvard.edu/abs/2019MNRAS.484.1401R} {484, 1401}

\bibitem[\protect\citeauthoryear{{Richardson} \& {Fairbairn}}{{Richardson} \&
  {Fairbairn}}{2014}]{richardson14}
{Richardson} T.,  {Fairbairn} M.,  2014, \mn@doi [\mnras]
  {10.1093/mnras/stu691}, \href
  {http://adsabs.harvard.edu/abs/2014MNRAS.441.1584R} {441, 1584}

\bibitem[\protect\citeauthoryear{Schneider, Trujillo-Gomez, Papastergis, Reed
  \& Lake}{Schneider et~al.}{2017}]{scheider17}
Schneider A.,  Trujillo-Gomez S.,  Papastergis E.,  Reed D.,   Lake G.,  2017,
  \mn@doi [Monthly Notices of the Royal Astronomical Society]
  {10.1093/mnras/stx1294}, 470, 1542

\bibitem[\protect\citeauthoryear{{Schwarzschild}}{{Schwarzschild}}{1979}]{Schwarzschild79}
{Schwarzschild} M.,  1979, \mn@doi [\apj] {10.1086/157282}, \href
  {https://ui.adsabs.harvard.edu/abs/1979ApJ...232..236S} {232, 236}

\bibitem[\protect\citeauthoryear{Sheather}{Sheather}{2004}]{Sheather04}
Sheather S.~J.,  2004, Statistical Science, 19, 588

\bibitem[\protect\citeauthoryear{Simon, Bolatto, Leroy, Blitz  \& Gates}{Simon
  et~al.}{2005}]{simon05}
Simon J.~D.,  Bolatto A.~D.,  Leroy A.,  Blitz L.,   Gates E.~L.,  2005,
  \mn@doi [The Astrophysical Journal] {10.1086/427684}, 621, 757

\bibitem[\protect\citeauthoryear{Spergel \& Steinhardt}{Spergel \&
  Steinhardt}{2000}]{spergel00}
Spergel D.~N.,  Steinhardt P.~J.,  2000, \mn@doi [Phys. Rev. Lett.]
  {10.1103/PhysRevLett.84.3760}, 84, 3760

\bibitem[\protect\citeauthoryear{{Stinson}, {Seth}, {Katz}, {Wadsley},
  {Governato}  \& {Quinn}}{{Stinson} et~al.}{2006}]{stinson06}
{Stinson} G.,  {Seth} A.,  {Katz} N.,  {Wadsley} J.,  {Governato} F.,   {Quinn}
  T.,  2006, \mn@doi [\mnras] {10.1111/j.1365-2966.2006.11097.x}, \href
  {http://adsabs.harvard.edu/abs/2006MNRAS.373.1074S} {373, 1074}

\bibitem[\protect\citeauthoryear{{Tollet} et~al.,}{{Tollet}
  et~al.}{2016}]{tollet15}
{Tollet} E.,  et~al., 2016, \mn@doi [\mnras] {10.1093/mnras/stv2856}, \href
  {https://ui.adsabs.harvard.edu/abs/2016MNRAS.456.3542T} {456, 3542}

\bibitem[\protect\citeauthoryear{{Walker} \& {Pe{\~n}arrubia}}{{Walker} \&
  {Pe{\~n}arrubia}}{2011}]{Walker11}
{Walker} M.~G.,  {Pe{\~n}arrubia} J.,  2011, \mn@doi [\apj]
  {10.1088/0004-637X/742/1/20}, \href
  {https://ui.adsabs.harvard.edu/abs/2011ApJ...742...20W} {742, 20}

\bibitem[\protect\citeauthoryear{{Walker}, {Mateo}  \& {Olszewski}}{{Walker}
  et~al.}{2009}]{Walker09a}
{Walker} M.~G.,  {Mateo} M.,   {Olszewski} E.~W.,  2009, \mn@doi [\aj]
  {10.1088/0004-6256/137/2/3100}, \href
  {https://ui.adsabs.harvard.edu/abs/2009AJ....137.3100W} {137, 3100}

\bibitem[\protect\citeauthoryear{{Wang}, {Dutton}, {Stinson}, {Macci{\`o}},
  {Penzo}, {Kang}, {Keller}  \& {Wadsley}}{{Wang} et~al.}{2015}]{wang15}
{Wang} L.,  {Dutton} A.~A.,  {Stinson} G.~S.,  {Macci{\`o}} A.~V.,  {Penzo} C.,
   {Kang} X.,  {Keller} B.~W.,   {Wadsley} J.,  2015, \mn@doi [\mnras]
  {10.1093/mnras/stv1937}, \href
  {http://adsabs.harvard.edu/abs/2015MNRAS.454...83W} {454, 83}

\bibitem[\protect\citeauthoryear{{Zhu}, {van de Ven}, {Watkins}  \&
  {Posti}}{{Zhu} et~al.}{2016}]{Zhu16}
{Zhu} L.,  {van de Ven} G.,  {Watkins} L.~L.,   {Posti} L.,  2016, \mn@doi
  [\mnras] {10.1093/mnras/stw2081}, \href
  {https://ui.adsabs.harvard.edu/abs/2016MNRAS.463.1117Z} {463, 1117}

\bibitem[\protect\citeauthoryear{{de Blok}, {Walter}, {Brinks}, {Trachternach},
  {Oh}  \& {Kennicutt}}{{de Blok} et~al.}{2008}]{deblok08}
{de Blok} W.~J.~G.,  {Walter} F.,  {Brinks} E.,  {Trachternach} C.,  {Oh}
  S.-H.,   {Kennicutt} Jr. R.~C.,  2008, \mn@doi [\aj]
  {10.1088/0004-6256/136/6/2648}, \href
  {http://adsabs.harvard.edu/abs/2008AJ....136.2648D} {136, 2648}

\bibitem[\protect\citeauthoryear{{van den Bosch} \& {de Zeeuw}}{{van den Bosch}
  \& {de Zeeuw}}{2010}]{bosch10}
{van den Bosch} R. C.~E.,  {de Zeeuw} P.~T.,  2010, \mn@doi [\mnras]
  {10.1111/j.1365-2966.2009.15832.x}, \href
  {https://ui.adsabs.harvard.edu/abs/2010MNRAS.401.1770V} {401, 1770}

\bibitem[\protect\citeauthoryear{van~der Marel}{van~der Marel}{1994}]{Marel94}
van~der Marel R.~P.,  1994, \mn@doi [Monthly Notices of the Royal Astronomical
  Society] {10.1093/mnras/270.2.271}, 270, 271

\makeatother
\end{thebibliography}


\section*{Appendix A: Details on the neural network model}
\label{sec:appendix}

A neural network can be formally described as a trainable and flexible approximation of a model $\mathbb{M} : d \rightarrow t$. The networks maps an input data $d$ to a prediction $\bar{t}$ of the target $t$. This network is parameterized by a set of trainable weights and a set of hyperparameters. The weights are iteratively optimized during training to minimize a particular loss function, which provides a measure of how close the network prediction $\bar{t}$ is to the target $t$. 

In this work, we use convolutional neural networks (CNNs), a class of deep neural networks (DNNs), to construct a neural network in which the input data $d$ are the two PDFs described in Section \ref{sec:pdfs}, and the targets $t$ are the inner slopes of the galaxy subsets associated to those two PDFs. We then make a mixture density  convolutional neural network (MDCNN) by embedding a mixture density layer within the CNN as the last layer.

\subsection*{Deep neural networks}

Any neural network is conformed by a set of neuron layers, defined by the following function:

\begin{equation}
    f(\mathbf{x}) = g(\mathbf{W} \cdot \mathbf{x} + b),
\end{equation}

\noindent where $\mathbf{x}$ is the input of the layer, $\mathbf{W}$ the weight matrix (which each element being the weight of each element of the vector $\mathbf{x}$) and $\mathbf{b}$ is a vector called the bias parameter of the layer. g($\mathbf{z}$) is known as the activation function, which purpose is to break the linearity between the input and the output of the neuron.

A DNN is a neural network conformed by more than one neuron layer. The layers between the input layer (the layer that takes as inputs the input data of the neural network) and the output layer (the layer that gives as output the outputs of the neural network) are called hidden layers.

A feed-forward DNN is a DNN where the neuron layers are evaluated in sequence, passing information from layer to layer without recurrence, which means we can describe the output $\mathbf{h}^{(l)}$ of the l-th layer as

\begin{equation}
    \mathbf{h}^{(l)} = g(\mathbf{W}^{(l)} \cdot \mathbf{h}^{(l-1)} + \mathbf{b}^{(l)}).
    \label{eq:layer}
\end{equation}

\noindent The training of the model is done by optimizing the weight matrices $\mathbf{W}^{(l)}$. A model is trained on a set of input data $d$ for which the targets $t$ are known iteratively. In each iteration, the network performance (the similarity between the outputs $\bar{t}$ and the targets $t$) is evaluated using a loss function, and the weights are actualized to minimize that function by an optimization algorithm. When the loss function stops decreasing and converges to a certain value, the network is said to be optimized. The performance evaluation is done, then, on a set of independent data the model has not seen during training.

\subsection*{Convolutional neural networks}

CNNs are a particular type of DNNs especially suited for problems where spatially correlated information is crucial. The main feature of a CNN is the presence of convolutional layers, which are constructed in a way that restrict neurons in one layer to receive information only from within a small neighborhood of the previous layer. This allows neurons to extract simple features from subsets of the previous layer, forming higher-order features in subsequent layers.

A convolutional layer is designed as follows: A convolutional kernel, commonly referred to as a filter, of a given size, encoding a set of neurons, is applied to each pixel (in the case of 2D images as inputs) of the input image and its vicinity as it scans through the whole region. A given pixel in a specific layer is only a function of the pixels in the preceding layer which are enclosed within the window defined by the kernel, known as the receptive field of the layer. This yields a feature map which encodes high values in the pixels which match the pattern encoded in the weights and biases of the corresponding neurons in the convolutional kernel, which are optimized during training \citep{ramanah20a}.

A convolutional layer may be described as a linear operation with the discrete convolution implemented via matrix multiplication. In terms of equation \ref{eq:layer}:

\begin{equation}
    \mathbf{h}^{(l)}_j = g\left(\sum_{i \in M_j} \mathbf{h}^{(l-1)} \times k^{(l)}_{ij} + \mathbf{b}^{(l)}_j\right),
    \label{eq:convolution}
\end{equation}

\noindent where $k$ is the convolutional kernel (the filter) and $M_j$ is the receptive field of the neuron $j$. One convolutional layer could have multiple filters, which repeat this operation with different kernels, constructing many feature maps per layer, known as channels. 

The receptive field is usually defined by the dimensions of the filter, the stride and the existence or not of padding. The application of the filter can be described as a process of sliding it over the input image of the convolutional layer. We call \textit{stride} to the number and direction of pixels you move the filter at each step, and \textit{padding} to the addition of empty pixels around the edges with the purpose of alleviating information loss around the edges. 

Usually, a CNN is a series of pairs of convolutional layers followed by a pooling layer as a subsampling or dimensionality reduction step, a process which will reduce the initial input image to a compact representation of features. Then, that representation is reshaped as a vector, which is subsequently passed to a sequence of dense layers \citep{lecun15}. This design allows the neural network to autonomously extract meaningful spatial features from the input image. The stack of several convolutional layers builds an internal hierarchical representation of features encoding the most relevant information from the input image. Stacking subsequent convolutional layers naturally strengthens the sensitivity of the most internal layers to features on increasingly larger scales, because the size of the receptive field becomes larger as we go deeper in the CNN.

\subsection*{Mixture density neural networks}

A MDNN is a network with layers whose outputs follow a multi-dimension probability distribution, called mixture density layers. This layers take as inputs $n$ nodes, with $n$ being the number of parameters in the desired distribution, transform their values to respect the parameter constraints of the distribution and interpret them as those parameters to construct it. When used as the last layer of the network, it allows join optimization of the features from the DNN together with a bayesian posterior backend, combining the advantages of deep feature extraction with probabilistic representation of the results.

\subsection*{Details on the architecture}

The schematic view of the architecture used in this work can be seen in Fig. \ref{fig:architecture}

The convolutional sequences are constructed using pairs of convolutional and pooling layers, followed by a dropout layer. The pooling layers downsample their input along its spatial dimension using the Max Pooling method, which takes the maximum value over a certain input window for each channel. 
The dropout layers randomly set input units to 0 with a certain frequency called the dropout rate, and scales the rest such that the sum over all inputs is unchanged. This is done to prevent overfitting during training.
The joint sequence of dense layer stars with a normalization layer that applies batch normalization to the nodes coming from the previous convolutional sequences. This normalization maintains the mean of the output close to 0 and its standard deviation close to 1.
The final mixture density layer gives a probability distribution defined in the range of possible inner slopes for a given galaxy subset and transform our double channel CNN into a MDCNN. This probability distribution is understood as a posterior under the prior distribution of inner slopes with which the network has trained, which allows us to evaluate the uncertainty of the individual predictions of the model.

\label{lastpage}

\end{document}